\documentclass[leqno, 12pt]{article}

\usepackage{pstricks}
\usepackage{amssymb}
\usepackage{mathrsfs} 
\usepackage{amsmath}
\usepackage{amsthm}
\usepackage{amsfonts}
\usepackage{hyperref} 
\usepackage{verbatim} 
\usepackage{graphicx}
 
\usepackage{flafter}
\usepackage{pstricks}
\usepackage{graphicx}
\usepackage{subcaption, float}
\usepackage{epstopdf} 
\usepackage{pgf,tikz} 
\usetikzlibrary{shapes} 
\usetikzlibrary{arrows} 
\usepackage{float} 

\def\thesection{\arabic{section}}
\renewcommand{\theequation}{\thesection.\arabic{equation}}

\newtheorem{theorem}{Theorem}[section]
\newtheorem{lemma}[theorem]{Lemma}
\newtheorem{proposition}[theorem]{Proposition}

\newtheorem{fact}[theorem]{Fact}

\newtheorem{problem}[theorem]{Problem}

\theoremstyle{definition}   
\newtheorem{remark}[theorem]{Remark}

\evensidemargin\oddsidemargin

\newcommand{\eqnsection}{
\renewcommand{\theequation}{\thesection.\arabic{equation}}
    \makeatletter
    \csname  @addtoreset\endcsname{equation}{section}
    \makeatother}
\eqnsection

\def\r{{\mathbb R}}

\def\P{{\bf P}}
\def\E{{\bf E}}

\def\z{{\mathbb Z}}

\def\ee{\mathrm{e}}

\def\T{{\mathbb T}}

\begin{document}

\baselineskip=18pt
\setcounter{page}{1}


\vglue50pt

\centerline{\bf\Large The dual Derrida--Retaux conjecture}  

\bigskip
\bigskip

 \centerline{by}

\medskip

 \centerline{Xinxing Chen\footnote{\scriptsize School of Mathematical Sciences, Shanghai Jiaotong University, 200240 Shanghai, China, {\tt chenxinx@sjtu.edu.cn} $\;$Partially supported by NSFC grant No.~12271351.},   Yueyun Hu\footnote{\scriptsize LAGA, Universit\'e Paris XIII, France, {\tt yueyun@math.univ-paris13.fr} $\;$Partially supported by ANR LOCAL.}, 
and Zhan Shi\footnote{\scriptsize  AMSS,
Chinese Academy of Sciences,  China, {\tt shizhan@amss.ac.cn}}}

\bigskip
\bigskip

\bigskip
\centerline{\it Dedicated to the memory of Professor Francis Comets}

\bigskip
\bigskip

{\leftskip=1.5truecm \rightskip=1.5truecm \baselineskip=15pt \small

\noindent{\slshape\bfseries Summary.} We consider a recursive system $(X_n)$ which was introduced by Collet et al.~\cite{collet-eckmann-glaser-martin}) as a spin glass model, and later by Derrida, Hakim, and Vannimenus~\cite{derrida-hakim-vannimenus} and by Derrida and Retaux~\cite{derrida-retaux} as a simplified hierarchical renormalization model. The system $(X_n)$ is expected to possess highly nontrivial universalities at or near criticality. In the nearly supercritical regime, Derrida and Retaux~\cite{derrida-retaux} conjectured that the free energy of the system decays exponentially with exponent $(p-p_c)^{-\frac12}$ as $p \downarrow p_c$. We study the nearly subcritical regime ($p \uparrow p_c$) and aim at a dual version of the Derrida--Retaux conjecture; our main result states that as $n \to \infty$, both $\E(X_n)$ and $\P(X_n\neq 0)$ decay exponentially with exponent $(p_c-p)^{\frac12 +o(1)}$, where $o(1) \to 0$ as $p \uparrow p_c$. 

\bigskip


\bigskip


} 

\bigskip
\bigskip

\section{Introduction}
\label{s:intro}

Fix an integer $m\ge 2$. Let $X^*_0 >0$ be a random variable taking values in $ \{ 1, \, 2, \ldots\}$ with  $\P(X^*_0\ge 2)>0$. 
For any random variable $X$, we write ${\mathscr L}_X$ for its law. Let $X_0$ be a random variable with law 
$$
{\mathscr L}_{X_0}
=
(1-p) \, \delta_0 + p\, {\mathscr L}_{X_0^*},
$$

\noindent where $p \in [0, \, 1]$, and $\delta_0$ denotes the Dirac measure at $0$. 

Consider the following recurrence relation: for all $n\ge 0$,
\begin{equation}
    X_{n+1}
    =
    (X_{n,1} + \cdots + X_{n,m} -1)^+ ,
    \label{iteration}
\end{equation}

\noindent where $X_{n,i}$, $i\ge 1$, are independent copies of $X_n$, and  $x^+ := \max\{ x, \, 0\}$ for all $x\in \r$.

Note that the law of each random variable $X_n$ is completely determined by the law of $X_0$, and that $X_n$ is stochastically non-decreasing in $p$. By \eqref{iteration}, we have
$$
\E(X_{n+1}) \le \E(X_{n,1} + \cdots + X_{n,m}) = m \, \E(X_n) \, ;
$$

\noindent therefore
$$
F_\infty(p):=\lim_{n\to \infty} \! \downarrow \frac{\E(X_n)}{m^n} \in [0, \infty)
$$

\noindent exists (and is called the free energy). The following phase transition was established by Collet et al.~\cite{collet-eckmann-glaser-martin}:

\bigskip

\noindent {\bf Theorem A (\cite{collet-eckmann-glaser-martin}).} {\it Assume $\E(X^*_0 \, m^{X^*_0})<\infty$.  Let  
\begin{equation}
    p_c
    =
    p_c(X_0^*)
    :=
    \frac{1}{1+ \E\{ [(m-1)X_0^*-1] m^{X_0^*}\} }
    \in (0, \, 1)\, .
    \label{p_c}
\end{equation}

{\rm (i)} If $p>p_c$, then $F_\infty(p)>0$.

{\rm (ii)} If $p\le p_c$, then $  \E(X_n) \to 0$.}

\bigskip

 Alternatively we may define $p_c$ through the free energy $F_\infty$ by letting $p_c:= \inf\{p: F_\infty(p) >0\}$.  Then the assumption $\E(X_0^* \, m^{X_0^*}) <\infty$ is equivalent to saying that $p_c>0$. We also remark that under the assumption of Theorem A,  
\begin{equation}
    \label{criticalmanifold} 
    p=p_c \, \, \Leftrightarrow\,\, \E(m^{X_0}) = (m-1) \E(X_0 \, m^{X_0}).
\end{equation}

\noindent More generally, it was proved in Collet et al.~\cite{collet-eckmann-glaser-martin} that for any $n\ge 0$,
\begin{equation}
    \label{criticalmanifold_general} 
    p=p_c \, \, \Leftrightarrow\,\, \E(m^{X_n}) = (m-1) \E(X_n \, m^{X_n}).
\end{equation}

\noindent In the language of Collet et al.~\cite{collet-eckmann-glaser-martin}, the identities in \eqref{criticalmanifold} and \eqref{criticalmanifold_general} mean that $X_0$ lies on the critical manifold.

We say that the system is subcritical if $p<p_c$, critical if $p=p_c$, and supercritical if $p>p_c$. 

The recursion \eqref{iteration} was introduced by Collet et al.~\cite{collet-eckmann-glaser-martin} as a simplified version of a spin glass model, and later by Derrida, Hakim,  and Vannimenus~\cite{derrida-hakim-vannimenus} and Derrida and Retaux~\cite{derrida-retaux} as a simplified hierarchical renormalization model, in order to understand the depinning transition of a line in presence of strong disorder, see  Giacomin,   Lacoin,  and Toninelli~\cite{giacomin-lacoin-toninelli}, also  Berger, Giacomin and Lacoin~\cite{berger-giacomin-lacoin} for an infinite order transition of a copolymer model. It was studied from the point of view of iterations of random functions (Li and Rogers~\cite{li-rogers}, Jordan~\cite{jordan}), and appeared as a special case in the family of max-type recursive models analyzed in the seminal paper of Aldous and Bandyopadhyay~\cite{aldous-bandyopadhyay}.  The recursion \eqref{iteration} is also connected to  a parking scheme recently investigated by Goldschmidt and Przykucki~\cite{goldschmidt-przykucki}, Curien and H\'enard~\cite{curien-henard},    Contat and Curien~\cite{contat-curien}, and Aldous et al.~\cite{aldous-contat-curien-henard}. See \cite{yz_bnyz} for an extension to the case when $m$ is random, and  \cite{HMP, 4authors} for an exactly solvable version in continuous time.

We call a Derrida--Retaux system the process $(X_n)_{n\ge0}$ defined in \eqref{iteration}. In fact,  our motivation of studying \eqref{iteration}  comes from Derrida and Retaux~\cite{derrida-retaux} who conjectured that $(X_n)_{n\ge0}$ possesses highly nontrivial universalities at or near criticality; these universality properties are believed to hold for a general class of pinning and hierarchical renormalization models, though none of them has been completely proved so far.  

In the nearly supercritical regime, the Derrida--Retaux conjecture (see \cite{derrida-retaux}) says that if $p_c>0$ (and  under some additional integrability conditions on $X^*_0$), then
\begin{equation}
     F_\infty(p)
     =
     \exp \Big( - \frac{C_1+o(1)}{(p-p_c)^{1/2}} \Big),
     \qquad
     p\downarrow p_c\, ,
     \label{conj:bernard}
\end{equation}

\noindent for some constant $C_1\in (0, \, \infty)$ depending on the law of $X_0^*$.  

A partial answer to the conjecture \eqref{conj:bernard} was given in \cite{bmvxyz_conjecture_DR}:  Assuming $\E[{X_0^*}^3 \, m^{X_0^*}] <\infty$, we have
 $$
   F_\infty (p)
   =
    \exp \Big( - \frac{1}{(p-p_c)^{\frac12 + o(1)}} \Big) ,
   \qquad
   p \downarrow p_c\, .
 $$
 
 \noindent In words, $\frac12$ was shown to be the correct exponent, in the exponential scale, for the free energy in  the nearly supercritical case.

For the critical regime $p=p_c$, we refer to \cite{bmxyz_questions} for a list of open questions concerning the behaviors of $X_n$, and to \cite{{xyz_sustainability}} for some recent progress. To insist on the criticality, we denote by $(Y_n)$ the  Derrida--Retaux system defined by the recurrence relation \eqref{iteration}, exactly in the same way as for $(X_n)$, but with \begin{equation}\label{Y_0}
{\mathscr L}_{Y_0}=(1-p_c)\,\delta_0+p_c \, {\mathscr L}_{X_0^*}.
\end{equation}

\noindent It was shown in  \cite{{xyz_sustainability}} that assuming the condition
\begin{equation}
    \label{exp-assumption} 
    \E(s^{X^*_0})<\infty, \qquad  \mbox{ for some $s>m$},
\end{equation}

\noindent we have 
$$
\E(Y_n) =  n^{-2+o(1)} \qquad  \mbox{and} \qquad \P(Y_n \ge 1)=  n^{-2+o(1)}, \qquad n \to \infty\, .
$$

In this paper, we study the subcritical regime. Assume from now on  that $p\in (0, \, p_c)$. There should be a kind of dual phenomenon for $p\uparrow p_c$ in the sense that a certain transition should also be expected   as in the Derrida--Retaux conjecture for the supercritical regime, and with the same exponent $\frac12$. More precisely, under some additional integrability conditions on $X_0^*$, we expect the existence of a constant $\kappa(p) \in (0, \, \infty)$ for all $p<p_c$ such that 
\begin{equation}
    \E(X_n) = \ee^{-(\kappa(p)+o(1))n}, \qquad n\to \infty,
    \label{exponent}
\end{equation}

\noindent where the exponent $\kappa(p)$ would satisfy
\begin{equation}
    \kappa(p) = (p_c-p)^{\frac12 +o(1)}, \qquad p \uparrow p_c \, .
    \label{dual_DR}
\end{equation}

We call \eqref{exponent} and \eqref{dual_DR} the dual Derrida--Retaux conjecture.  Roughly saying,  the critical manifold, characterised by the identity on the right-hand-side of \eqref{criticalmanifold}, says that if the initial distribution lies on the critical manifold, then the system always lies on the critical manifold. If the initial distribution does not exactly lie on the critical manifold, but only in a neighbourhood, with distance $\varepsilon$ (by Collet et al.~\cite{collet-eckmann-glaser-martin}, this is equivalent to saying that $|p-p_c|$ is of order $\varepsilon$), then for a long time, of order $\varepsilon^{-\frac{1}{2}}$, the system lies in the $\varepsilon$-neighbourhood of the critical manifold before drifting away definitely. This phenomenon does not depend on the sign of $p-p_c$; in other words, it is common for both supercritical and subcritical regimes. The quantity of time $\varepsilon^{-\frac{1}{2}}$ leads to the Derrida--Retaux conjecture in the supercritical case, and to the dual conjecture in the subcritical case. This is why we would also expect to see the exponent $\frac12$ in the dual conjecture for the subcritical regime, as in the Derrida--Retaux conjecture for the nearly supercritical regime.

Unfortunately, we have not been able to show the existence of $\kappa(p)$ in \eqref{exponent}. The main result of this paper is a weaker form of \eqref{exponent} and \eqref{dual_DR}, which confirms that the exponent $\frac12$ does appear in the subcritical case:

\medskip

\begin{theorem} 
 \label{t:main} 
 
 Assume \eqref{exp-assumption}.  Let $p= p_c - \varepsilon$ with $\varepsilon \in (0, \, p_c)$. 

{\rm (i)} There exists some positive constant $C_2$, independent of $\varepsilon$, such that    
\begin{equation}
    \label{upper} 
    \limsup_{n\to\infty} \frac1n \log \E(X_n)
    \le 
    - C_2 \, \varepsilon^{\frac12}.
\end{equation}

{\rm (ii)} We have  
\begin{equation}
    \label{lower} 
    \liminf_{n\to\infty} \frac1n \log \E(X_n)
    \ge 
    - \varepsilon^{\frac12+o(1)},
\end{equation}

\noindent where $o(1)$ denotes some quantity which tends to $0$ as $\varepsilon\to0$. 

\end{theorem}
  
\medskip

We mention that the upper bound \eqref{upper} holds for $\E (m^{X_n})-1$ in place of $\E(X_n)$, see Proposition \ref{p:ub}, whereas the lower bound \eqref{lower} holds for $\P(X_n \ge 1)$ in place of $\E(X_n)$, see Proposition \ref{p:lb}.   
  
\medskip

\begin{remark} 
\label{r:counter-example}

 In \cite{bmvxyz_conjecture_DR}, the optimal condition for the validity of the usual Derrida--Retaux conjecture was proved to be $\E((X_0^*)^3 m^{X_0^*})<\infty$. For the dual conjecture, we claim that \eqref{exp-assumption} is optimal.

To see why \eqref{exp-assumption} is also necessary for the validity of Theorem \ref{t:main}, we first remark that by definition, for all $n\ge 1$, 
$$
X_n \ge \sum_{i=1}^{m^n} {\bf 1}_{\{ X_{0,i} \ge n+1\} } \, ,  $$

\noindent where $X_{0,i}$, $i\ge 1$, denote as before independent copies of $X_0$. Then  
\begin{equation*}
    \E(X_n) \ge m^n \, \P(X_0 \ge n+1)= p \, m^n \, \P(X^*_0 \ge n+1).
\end{equation*}

 Assume that \eqref{upper} holds for $p= p_c-\varepsilon$ with $\varepsilon\in (0, \, p_c)$.  We have, for some constant $C_3>0$ and all sufficiently large $n$,
$$
\E(X_n) \le \ee^{-C_3 n} \, .
$$

\noindent 
It follows that for all sufficiently large $n$, 
$$
\P(X_0^* \ge n+1) 
\le
\frac{m^{-n}}{p_c- \varepsilon} \, \E(X_n)
\le
\frac{m^{-n}}{p_c- \varepsilon} \, \ee^{-C_3 n} \, .
$$

\noindent Consequently, if \eqref{upper} holds for $\varepsilon \in (0, \, p_c)$, then $\E(s^{X_0^*}) <\infty$ as long as $s\in (0, \, m \ee^{C_3})$. This shows the optimality of \eqref{exp-assumption}. \qed
\end{remark}

\medskip

The proof of Theorem \ref{t:main} relies on the aforementioned intuitive ideas, by using a coupling between the subcritical system $(X_n)$ and the critical Derrida--Retaux system $(Y_n)$ (see \eqref{Y_0}).  While the upper bound \eqref{upper} in Theorem \ref{t:main} follows from  some precise estimates on the generating functions of $X_n$, the lower bound \eqref{lower} relies on a coupling argument for the ``survival probability" $\P(X_n \ge 1)$ for subcritical systems and the Laplace transform of the number of open paths for critical systems. More precisely, let $N_n$ (defined in \eqref{def-N}) denote the number of open paths. Then for $p\in (0, \, p_c)$ and $n\ge 1$, 
$$
\P (X_n \ge 1) 
\ge 
\E\Big[ \Big( \frac{p}{p_c} \Big)^{\! N_n} \, {\bf 1}_{\{ Y_n \ge 1\}}\Big] ;
$$

\noindent see Theorem \ref{t:coupling}. Under the conditional probability $\P( \, \bullet \, | \, Y_n \ge 1)$, $N_n$ is typically of order $n^2$, but a small deviation result (see \eqref{smallN}) says that for as $n\to \infty$, 
$$
\P (1\le N_n \le j n) \ge \exp \Big( - \frac{n}{j^{1+o(1)}} \Big) ,
$$

\noindent with $o(1)$ denoting an expression that does not depend on $n$ and that converges to $0$ when $j \to\infty$. Taking $j$ to be the integer part of $\varepsilon^{-1/2}$, we obtain the lower bound \eqref{lower}. 

\bigskip

The rest of the paper is organised as follows. 
The upper and lower bounds in Theorem \ref{t:main} are proved in Sections \ref{s:ub} and \ref{s:lb}, respectively. Some further remarks and questions are presented in Section \ref{s:final}. 

Throughout the paper, $C_i$, $1\le i \le 22$, denote some positive constants whose values do not depend on $p$.

\section{Upper bound}
\label{s:ub}

The proof of the upper bound in Theorem \ref{t:main} is purely analytical, based on study of the moment generating function of the system. By the monotonicity in $p$ of $\E(X_n)$, it suffices to prove the upper bound \eqref{upper} in Theorem \ref{t:main} for sufficiently small $\varepsilon$. The aim of this section is to prove the following result.

\medskip

\begin{proposition}
 \label{p:ub}

 Under \eqref{exp-assumption}, there exist a positive constant $C_2$ and a sufficiently small $\varepsilon_0 \in (0, \, \frac{p_c}2)$ such that for all $p=p_c-\varepsilon$ with $\varepsilon \in (0, \, \varepsilon_0)$, there exists some $s_0>m$ such that  \begin{equation}
     \label{upper1} 
     \limsup_{n\to\infty} \frac1{n} \log \big(\E(s_0^{X_n})-1\big) 
     \le 
     - C_2 \, \varepsilon^{1/2}.
\end{equation}

\end{proposition}

We note that $s_0$ in \eqref{upper1} can be chosen such that  $s_0-m$ is of order $\varepsilon^{1/2}$ as $\varepsilon\to0$. 

Since $\E(X_n)\le \frac{1}{m-1}(\E(m^{X_n})-1) \le \frac{1}{m-1}(\E(s_0^{X_n})-1)$,  the upper bound \eqref{upper} in Theorem \ref{t:main} will follow immediately from \eqref{upper1}.

The rest of the section is devoted to the proof of Proposition \ref{p:ub}. For the sake of clarity, the proof is divided into two parts. The first part collects some known estimates of the moment generating function for subcritical systems, followed by the second part which contains the proof of Proposition \ref{p:ub}.

\subsection{Preliminaries on the moment generating function}

For any $n\ge 0$, we write
\begin{equation}
    H_n(s) := \E(s^{X_n}) \, .
    \label{H}
\end{equation}

\noindent In order to insist on the criticality, we write
$$
G_n(s) := \E(s^{Y_n}) \, ,
$$

\noindent where $(Y_n)$ denotes as before the critical Derrida--Retaux system. 
We first list some known facts about {\it critical} systems.

\medskip

\begin{fact}
\label{f:collet_et_al}

 {\bf (The moment generating function at criticality).}
 Assume $\E({X_0^*}^3 m^{X^*_0})<\infty$. We have 
 \begin{eqnarray}
     \sup_{n\ge0} G_n(m)
  &\le& m^{\frac{1}{m-1}}, 
     \label{supGn}
     \\
     C_4\, n^2 \le \prod_{i=0}^{n-1} G_i(m)^{m-1} 
  &\le& C_5 \, n^2, 
     \qquad 
     \forall n\ge 1,
     \label{prodGi} 
 \end{eqnarray} 

 \noindent for some positive constants $C_4$ and $C_5$. 
\end{fact}

\medskip

Fact \ref{f:collet_et_al} borrows from \cite{collet-eckmann-glaser-martin} in case $m=2$. For general $m\ge 2$, see \cite[Lemma 3]{bmxyz_questions} for \eqref{supGn} and \cite[Propositions 1 and 2]{bmxyz_questions} for \eqref{prodGi}. 

We consider from now on the {\it subcritical} regime $p\in (0, \, p_c)$. Plainly $X_n$ is stochastically smaller than $Y_n$ for all $n\ge 0$. In particular, $H_n(s)\le G_n(s)$ for any $s\in [0, m]$. 

Define 
\begin{equation}
    \label{delta_n}
    \delta_n:=H_n(m)-m(m-1)H_n'(m), \qquad n\ge 0. 
\end{equation}
 
\noindent Note that $\delta_n= \E[ (1-(m-1)X_n) m^{X_n}] \le \P(X_n=0)\le 1$. We shall see that $\delta_n$ is in fact nonnegative.

Let $\varepsilon := p_c-p>0$. From $\E(m^{Y_0})=(m-1)\E(Y_0m^{Y_0})$ (see \eqref{criticalmanifold_general}), we deduce that
\begin{equation}
    \delta_0 = C_6\, \varepsilon,
    \label{delta0}
\end{equation}

\noindent where $C_6=\E[((m-1)X_0^*-1)m^{X_0^*}]+1$. 

\medskip

\begin{lemma}\label{l:recu} 

 {\bf (The subcritical moment generating function).}
Suppose $\E({X_0^*}^3 m^{X^*_0})<\infty$ and $p \in (0, \, p_c)$. Then for all $n\ge1$,  
\begin{eqnarray}
    \label{e:iterdelta}
    \delta_n
 &=& \delta_0\prod_{i=0}^{n-1} H_i(m)^{m-1} \in (0, 1],
    \\
    \prod_{i=0}^\infty H_i(m)^{m-1}
 &\le& \frac1{\delta_0} , 
    \label{e:prod}
    \\
    \delta_n 
 &\le&  C_5 \, n^2\delta_0, 
    \label{delta_n<}
    \\
    C_7 \, \E(X_n^3 m^{X_n}) 
 &\le& \prod_{i=0}^{n-1} H_i(m)^{m-1} ,  
    \label{e:xn3}
\end{eqnarray}  

\noindent where $C_5$ is as in \eqref{prodGi}, and $C_7$ is some positive constant, independent of $p$. 

\end{lemma}

\medskip

\noindent {\it Proof.} The idea of the proof, based on the iteration of a suitable combination of the generating function $H_n$ and its derivatives (up to the third derivative), goes back to \cite{collet-eckmann-glaser-martin, derrida-retaux} and has already been explored in \cite{bmxyz_questions, xz_stable}; in particular, \eqref{e:xn3} is essentially Lemma 1 of \cite{xz_stable}. We give here the details for the sake of completeness. Let $s\in [0, \, m]$ and $n\ge 0$. By \eqref{iteration},  
$$
H_{n+1}(s) = \frac1s \, H_n(s)^m + (1-\frac1s)H_n(0)^m  .
$$

\noindent Taking the derivative with respect to $s$ gives that 
$$
    H_{n+1}'(s)
    = \frac{m}{s}\, H_n'(s) \, H_n(s)^{m-1} - \frac{1}{s^2} \, H_n(s)^m + \frac{1}{s^2} \, H_n(0)^m \, ,
$$

\noindent from which it follows that
$$
    (s-1)s\, H_{n+1}'(s) - H_{n+1}(s)
    =
    [ m(s-1) \, H_n'(s) - H_n(s) ] \, H_n(s)^{m-1} \, ,
$$

\noindent Taking $s=m$, the identity reads: 
$$
    \delta_{n+1}= \delta_n \, H_n(m)^{m-1}
     \, ,
$$

\noindent which yields the equality in \eqref{e:iterdelta}. Recall that $\delta_0>0$ for $p<p_c$. Thus $\delta_n>0$. Since we have already observed that $\delta_n \le 1$, this yields \eqref{e:iterdelta}. Letting $n\to\infty$ gives \eqref{e:prod}, whereas \eqref{delta_n<} is a consequence of \eqref{prodGi} because $H_i(m) \le G_i(m)$ for any $i\ge0$. 

To show \eqref{e:xn3},  further differentiations lead to 
 \begin{eqnarray}
H''_{n+1}(s) + \frac2{s} H'_{n+1}(s)
&=&
\frac{m}{s} H''_n(s) H_n(s)^{m-1} + \frac{m(m-1)}{s}  H'_n(s)^2\,  H_n(s)^{m-2}, \label{derivative2}
\\
s H'''_{n+1}(s) + 3 H''_{n+1}(s)
&=&
m H'''_n(s) H_n(s)^{m-1} + 3 m (m-1) H'_n(s) H''_n(s)   H_n(s)^{m-2}  \nonumber
\\
&& \quad + m(m-1)(m-2) H'_n(s)^3 H_n(s)^{m-3}. \label{derivative3}
\end{eqnarray}

  Define  for all $n\ge0$, $$ \mathcal{D}_n(m)
    :=
    (m-1) ( m H'''_n(m)+ 3 H''_n(m)) + (m-2) (H''_n(m) + \frac2{m} H'_n(m)) .$$

\noindent  Taking $s=m$ in \eqref{derivative2} and \eqref{derivative3}, we get that 
\begin{eqnarray*}  
    \mathcal{D}_{n+1}(m)
 &=& H_n(m)^{m-1} \mathcal{D}_n(m) - 3 (m-1) \delta_n H''_n(m) H_n(m)^{m-2} 
    \\
 && \hskip-50pt
    - \frac{m-2}{m} H'_n(m) H_n(m)^{m-3} [ 2 H_n(m)^2 - m^2(m-1)^2 H'_n(m)^2- m(m-1) H'_n(m) H_n(m)].
\end{eqnarray*}

\noindent Note that $H_n(m)>  m(m-1) H'_n(m)$, so the $[\cdots]$ term, $2 H_n(m)^2 - m^2(m-1)^2 H'_n(m)^2- m(m-1) H'_n(m)H_n(m)$, is greater than $H_n(m)^2 - m(m-1) H'_n(m)H_n(m) = H_n(m) \delta_n$, which is positive. Therefore, for all $n\ge 0$,
$$
    \mathcal{D}_{n+1}(m)
    \le
     H_n(m)^{m-1}\, \mathcal{D}_n(m).
$$

\noindent Iterating the inequality gives that
$$
\mathcal{D}_n(m)
\le 
\mathcal{D}_0(m)\prod_{i=0}^{n-1}H_i(m)^{m-1}, 
\qquad 
\forall\, n\ge 1.
$$

\noindent By definition, $\mathcal{D}_n(m) \ge m(m-1) H_n'''(m)$; thus
$$
m(m-1) H_n'''(m)
\le
\mathcal{D}_0(m)\prod_{i=0}^{n-1}H_i(m)^{m-1}.
$$ 

\noindent   Note that for any $k\ge 0$, $k^3 m^{k} \le \frac92 k (k-1)(k-2) m^k + 8m^2 \, {\bf 1}_{\{k\le 2\}}$, it follows that $$ \E(X_n^3 m^{X_n}) \le \frac92 m^3  H_n'''(m) + 8m^2 \le ( \frac{9 m^2}{2(m-1)} \mathcal{D}_0(m)+ 8m^2)\prod_{i=0}^{n-1}H_i(m)^{m-1} .$$

 Since $X_0$ is stochastically smaller than $Y_0$, we have $\mathcal{D}_0(m) \le (m-1) (m G'''_0(m)+ 3 G''_0(m)) + (m-2) (G''_0(m) + \frac2m G'_0(m)) =: \mathcal{D}^Y_0(m)$. This yields \eqref{e:xn3} with $C_7 := (\frac{9m^2}{2(m-1)}  \mathcal{D}^Y_0(m)+ 8m^2)^{-1}$.\qed

\medskip

\begin{fact}
\label{l:ub_known_1}

 {\bf (\cite[Lemma 4.5]{xyz_sustainability})} Let $p\in (0, \, 1)$. If there exist $t>m$, $\theta \in (0,\, 1)$ and an integer $M\ge 0$ such that $\frac{m}{t}\, [\E(t^{X_M})]^{m-1} \le \theta$, then 
\begin{equation}
    \label{HM+n} 
    \E(t^{X_n}) \le 1 + (t-m) \theta^{n-M}, \qquad \forall\, n\ge M. \end{equation} 

\end{fact}

\begin{fact}
\label{l:ub_known_2} 

 {\bf (\cite[Corollary 1]{xz_stable})} Under \eqref{exp-assumption}, there exist positive constants $C_8$ and $C_9$ 
such that for all sufficiently large integer $n$,\footnote{Strictly speaking, we first consider a truncating version of $(X_n)$: For any integer $L \ge 1$, let $Z_0 = Z_0(L):= X_0 \, {\bf 1}_{\{X_0\le L\}}$ and $(Z_n)$ be the associated Derrida--Retaux system. As explained in Section 3 of \cite{xz_stable}, we may apply \cite[Corollary 1]{xz_stable} to $(Z_n)$ and get \eqref{CS-stableY} for $Z_M$ in place of $X_M$, and with positive constants $C_8$ and $C_9$ that are independent of $L$. Then we let $L \to \infty$ and obtain \eqref{CS-stableY}.}
\begin{equation}
    \E (Y_n^2 s^{Y_n})
    \le 
    C_8 \, n,
    \qquad 
    s := m+\frac{C_9}{n}. 
    \label{CS-stableY}
\end{equation}

\noindent A fortiori, for all $p\in (0, \, p_c]$,
\begin{equation}
    \E (X_n^2 s^{X_n})
    \le 
    C_8 \, n, 
    \qquad 
    \forall \, s\in [m, \, m+\frac{C_9}{n}]. 
    \label{CS-stable}
\end{equation}

\end{fact}
  
 \subsection{Proof of Proposition \ref{p:ub}}

Write, as before, $H_n(s) := \E(s^{X_n})$, $n\ge 0$, for the moment generating function of $X_n$.

\medskip

\begin{lemma}
 \label{l:prod_lower} 
 
 Suppose $\E({X^*_0}^3 m^{X^*_0})<\infty$. There exist constants $C_{10}>0$ and $C_{11}>0$, such that for $p=p_c-\varepsilon$ with $\varepsilon \in (0, \frac{p_c}2)$ and $1\le n\le C_{10} \, \varepsilon^{-\frac12}$,
$$
\prod_{i=0}^{n-1} H_i(m)^{m-1}\ge C_{11} \, n^2.
$$
\end{lemma}

\medskip

\noindent {\it Proof of Lemma \ref{l:prod_lower}.}  Write $b_n=\prod_{i=0}^{n-1} H_i(m)^{m-1}$. Note that  $1\le H_i(m)\le G_i(m)\le  m^{\frac1{m-1}}$ (see \eqref{supGn}). Let $C_{12}>0$ (whose value depends only on $m$) be such that  $(1+x)^{\frac{m-1}{2}}-1\ge C_{12} \, x$ for $x\in [0,m^{\frac1{m-1}}-1]$. Then 
$$
b_{n+1}^{\frac{1}{2}}-b_{n}^{\frac{1}{2}}=b_n^{\frac{1}{2}}(H_n(m)^{\frac{m-1}{2}}-1)\ge C_{12} \, b_n^{\frac{1}{2}} \, (H_n(m)-1).
$$

For the term on the right-hand side, we note that $b_n \ge C_7\,  \E(X_n^3 m^{X_n})$ by \eqref{e:xn3}, and that $H_n(m)-1\ge (1-\frac{1}{m})\E(m^{X_n} {\bf 1}_{\{X_n\ge 1\}})$. Since $\E(X_n^3m^{X_n}) \, [\E(m^{X_n} {\bf 1}_{\{X_n\ge 1\}})]^2 \ge [\E(X_n m^{X_n})]^3$ (H\"{o}lder's inequality), it follows that
\begin{eqnarray*}
    b_n^{\frac{1}{2}} \, (H_n(m)-1) 
 &\ge& C_7^{\frac12} \, (1-\frac1m)\, [\E(X_n^3 m^{X_n})]^{\frac12} \E(m^{X_n} {\bf 1}_{\{X_n\ge 1\}}) 
    \\
 &\ge& C_7^{\frac12} \, (1-\frac1m)\, [\E(X_n m^{X_n})]^{\frac32}. 
\end{eqnarray*}

\noindent Consequently, for all $n\ge 1$ and with $C_{13} := C_{12} \, C_7^{\frac12} \, (1-\frac{1}{m})$,
$$
b_{n+1}^{\frac{1}{2}}-b_{n}^{\frac{1}{2}}
\ge 
C_{13} \, [\E(X_n m^{X_n})]^{\frac32}.
$$

\noindent By definition, $\delta_n = H_n(m)-m(m-1)H_n'(m)$, thus $\E(X_n m^{X_n}) = \frac{H_n(m)-\delta_n}{m-1} \ge \frac{1-\delta_n}{m-1}$. Since $\delta_n\le C_5 \, n^2\delta_0 =  C_5 C_6\, n^2 \varepsilon$ (see \eqref{delta_n<} for the inequality, and \eqref{delta0} for the equality), we have $\delta_n\le \frac12$ for $n\le (2C_5 C_6 \varepsilon)^{-\frac12}$. Therefore,
$$
\E(X_n m^{X_n}) \ge \frac{1-\frac12}{m-1} = \frac{1}{2(m-1)}, \qquad \forall\, 1\le n\le (2 C_5 C_6 \varepsilon)^{-\frac12}.
$$

\noindent As such, for all $1\le n\le (2 C_5 C_6 \varepsilon)^{-1/2}$, we have $b_{n+1}^{1/2}-b_n^{1/2} \ge \frac{C_{13}}{[2(m-1)]^{3/2}}$, which implies that $b_{n+1}^{1/2} \ge \frac{C_{13}}{[2(m-1)]^{3/2}} \, n +1$. 
The lemma follows immediately.\qed

\medskip

\begin{lemma}
 \label{l:momentupper} 
 
 Assume \eqref{exp-assumption}. For any sufficiently small $\varepsilon_0>0$,  there exist positive constants $C_{14}$, $C_{15}$, $C_{16}$ and $C_{17}$ such that for any $p=p_c-\varepsilon$ with $\varepsilon\in (0, \, \varepsilon_0)$, we can find some an integer $M\in [C_{14}\, \varepsilon^{-\frac12}, \, C_{15}\, \varepsilon^{-\frac12}]$ and some $s_* \in (m, \, m+\frac{C_{16}}{M}]$ such that   
$$ 
\frac{m}{s_*}\, H_M (s_*)^{m-1}< 1-C_{17}\, \varepsilon^{\frac12}.
$$

\end{lemma}

\medskip
\noindent {\it Proof.} Let $\varepsilon_0\in (0, \, \frac{p_c}2)$ be small (how small will be determined later) and $\varepsilon \in (0, \, \varepsilon_0)$. By Lemma \ref{l:prod_lower}, we have\footnote{For notational brevity, we write $C_{10}\, \varepsilon^{-\frac12}$ instead of $\lfloor C_{10}\, \varepsilon^{-\frac12}\rfloor$. Similar simplifications apply elsewhere.}
\begin{equation}
    \prod_{i=0}^{C_{10}\varepsilon^{-\frac12} -1} H_i(m)^{m-1}
    \ge 
    C_{11} \, (C_{10}\varepsilon^{-\frac12} -1)^2 
    \ge 
    \frac{C_{18}}{\varepsilon},
    \label{pf_l:momentupper_1}
\end{equation}

\noindent where $C_{18}:= \frac14 C_{11} C^2_{10}$; here appears the first constraint on $\varepsilon_0$: $\varepsilon_0$ is sufficiently small such that $C_{10}\, \varepsilon_0^{-\frac12}\ge 2$ and that $C_{11} \, (C_{10}\varepsilon^{-\frac12} -1)^2 \ge \frac{C_{18}}{\varepsilon}$ for $\varepsilon\in (0, \, \varepsilon_0)$. By \eqref{e:prod}, we have $\prod_{i=0}^\infty H_i(m)^{m-1}\le \frac1{\delta_0}$ which is $\frac{1}{C_6\, \varepsilon}$ (see \eqref{delta0}). Thus
$$
\prod_{i=C_{10}\varepsilon^{-\frac12}}^\infty H_i(m)^{m-1}
\le 
\frac{1}{C_6\, C_{18}} ; 
$$ 

\noindent in other words, 
\begin{equation}
    \prod_{i=C_{10}\varepsilon^{-\frac12} }^\infty H_i(m) \le \big(\frac{1}{C_6\, C_{18}}\big)^{\frac1{m-1}}=:C_{19} \, .
    \label{e:comparison_M}
\end{equation}

Let $\lambda>0$ be a small constant whose value will be given later. Let $C_{15} := C_{10}\, \ee^{\frac{2 C_{19}}{\lambda}}$. We have
$$
\prod_{i=C_{10}\varepsilon^{-\frac12} }^{C_{15}\varepsilon^{-\frac12} } (1+\frac{\lambda}{i})
\ge 
\sum_{i=C_{10}\varepsilon^{-\frac12} }^{C_{15}\varepsilon^{-\frac12} } \frac{\lambda}{i}
\ge 
\lambda\, \log \frac{C_{15}}{C_{10}} 
=
2 \, C_{19}.
$$

\noindent Compared with \eqref{e:comparison_M}, it follows that there exists $M\in  [C_{10}\varepsilon^{-\frac12}, C_{15}\varepsilon^{-\frac12}] \cap \z$ such that
\begin{equation}
    \label{e:H_M_upper}
    H_M(m) < 1+\frac{\lambda}{M}. 
\end{equation}

On the other hand, since $M \ge C_{10}\varepsilon^{-\frac12}$, we have 
$$
    \prod_{i=0}^{M-1}H_i(m)^{m-1} 
    \ge 
    \prod_{i=0}^{C_{10}\varepsilon^{-\frac12} -1} H_i(m)^{m-1} 
    \ge 
    \frac{C_{18}}\varepsilon \, ,
$$

\noindent the second inequality being from \eqref{pf_l:momentupper_1}. Since $H_M(m)-m(m-1)H_M'(m)= \delta_M$, which equals $\delta_0 \prod_{i=0}^{M-1}H_i(m)^{m-1}$ (see \eqref{e:iterdelta}), and $\delta_0 = C_6 \, \varepsilon$ (see \eqref{delta0}), it follows that 
$$
H_M(m)-m(m-1)H_M'(m) 
\ge 
C_6 \, \varepsilon \, \frac{C_{18}}\varepsilon
=
C_6 \, C_{18}
=
2 C_{20},
$$

\noindent with $C_{20} := \frac{C_6 \, C_{18}}{2}$. 

By assumption \eqref{exp-assumption}, there exists some $s_1>m$,  independent of $p$,  such that $\E (s_1^{X_0}) < \infty$. Since $H_M-m(m-1)H_M'$ is $C^\infty$ on $[0, \, s_1)$, we can apply the mean-value theorem to see that for all $s \in (m, \, s_1)$, there exists $u\in [m, \, s]$ such that
\begin{eqnarray}
 &&H_M(s)-m(m-1)H_M'(s)
    \nonumber
    \\
 &=& H_M(m)-m(m-1)H_M'(m)+(s-m)[H_M'(u)-m(m-1)H_M''(u)]
    \nonumber
    \\
 &\ge& 2 C_{20}-(s-m)m(m-1)H_M''(u).
    \label{H_M-m(m-1)H_M_prime}
\end{eqnarray}

\noindent Let $C_{21}:=\min\{\frac{C_{20}}{m(m-1)C_8}, C_9, \, \frac{s_1-m}{2}\}$ (where $C_8$ and $C_9$ are the positive constants in Fact \ref{l:ub_known_2}). We apply \eqref{H_M-m(m-1)H_M_prime} to $s=s_*:= m+\frac{C_{21}}{M} \in (m, \, s_1)$; then $(s_*-m)m(m-1)H_M''(u) \le \frac{C_{21}}{M}\, m(m-1) \, C_8 M$ (by \eqref{CS-stable}), so that
\begin{equation}
    H_M(s_*)-m(m-1)H_M'(s_*)
    \ge 
    2C_{20} - C_8C_{21} m(m-1)
    \ge
    C_{20} \, .
    \label{H_M-m(m-1)H_M_prime_bis}
\end{equation}

\noindent On the other hand, by the convexity of $H_M(\cdot)$, 
$$
H_M'(s_*)
\ge 
\frac{H_M(s_*)-H_M(m)}{s_*-m} 
=
\frac{M}{C_{21}} (H_M(s_*)-H_M(m)) 
\ge 
\frac{M}{C_{21}} (H_M(s_*)-1- \frac{\lambda}{M}),
$$

\noindent where the last inequality follows from \eqref{e:H_M_upper}. Going back to \eqref{H_M-m(m-1)H_M_prime_bis}, we obtain that
$$
H_M(s_*)
\ge
 C_{20}
+
m(m-1) \, \frac{M}{C_{21}} (H_M(s_*)-1-\frac{\lambda}{M}) .
$$

\noindent Let $C_{22} := \frac{m(m-1)}{C_{21}}$. We impose a new constraint on $\varepsilon_0$: $C_{10}\varepsilon_0^{-\frac12} > \frac1{C_{22}}$. Then $C_{22}M >1$ as long as $\varepsilon \in (0, \, \varepsilon_0)$. We get that
$$
H_M(s_*)-1
\le
\frac{1- C_{20} +\lambda C_{22}}{C_{22} M-1} \, .
$$

\noindent Now, we choose $\lambda := \frac{C_{20}}{2C_{22}}$; thus
$$
\frac{m}{s_*}\, H_M (s_*)^{m-1}
=
\frac{H_M (s_*)^{m-1}}{1+ \frac{C_{21}}{mM}}
\le
\frac{1}{1+ \frac{C_{21}}{mM}} \Big( 1+\frac{1- \frac12 C_{20}}{C_{22} M -1} \Big)^{\! m-1} \, .
$$

\noindent We further require that $\varepsilon_0$ is small enough so  that $\frac{1}{1+ \frac{C_{21}}{mM}} ( 1+\frac{1- \frac12 C_{20}}{C_{22} M -1})^{m-1} \le 1 - \frac{C_{20}C_{21}}{3mM}$ for all $M \ge C_{10}\varepsilon^{-\frac12} $ with $\varepsilon\in (0,\varepsilon_0)$. [This is possible because $-\frac{C_{21}}{m} + \frac{(1- \frac12 C_{20}) (m-1)}{C_{22}} < - \frac{C_{20}C_{21}}{3m}$.] Consequently, for $\varepsilon\in (0, \, \varepsilon_0) $,
$$
\frac{m}{s_*}\, H_{M}(s_*)^{m-1}
\le
1- \frac{C_{20}C_{21}}{3mM}
\le 
1 - C_{17} \, \varepsilon^{\frac12} \, ,
$$

\noindent where $C_{17}:=\frac{C_{20} C_{21}}{3m C_{10}}$. This yields Lemma \ref{l:momentupper}.\qed

\bigskip

We now have all the ingredients for the proof of Proposition \ref{p:ub}.

\bigskip

\noindent {\it Proof of Proposition \ref{p:ub}.} By Lemma \ref{l:momentupper}, there exist $\varepsilon_0>0$ and $C_{17}>0$ such that for $p=p_c-\varepsilon$ with $\varepsilon \in (0, \, \varepsilon_0)$, there exist $s_*>m$ and an integer $M$ satisfying $\frac{m}{s_*}\, [\E(s_*^{X_M})]^{m-1} < 1-C_{17}\, \varepsilon^{\frac12}$. So we are entitled to apply Fact \ref{l:ub_known_1} to $t:= s_*$ and $\theta := 1-C_{17}\, \varepsilon^{\frac12}$ to see that for all integer $n\ge M$, 
$$
\E(s_*^{X_n}) 
\le 
1+(s_*-m) (1-C_{17}\,\varepsilon^{\frac12})^{n-M}
\le
1+(s_*-m) \ee^{-(n-M) C_{17}\,\varepsilon^{\frac12}} \, .
$$

\noindent Therefore,
$$
\limsup_{n\to\infty} \frac1n \log \big(\E (s_*^{X_n})-1\big) 
\le 
- C_{17}\,\varepsilon^{\frac12} ,
$$

\noindent which yields the desired inequality \eqref{upper1}.\qed

\section{Lower bound}
\label{s:lb}

The lower bound in Theorem \ref{t:main} will be a straightforward consequence of the following result.

\medskip

\begin{proposition} 
 \label{p:lb} 
 
 Assume \eqref{exp-assumption}. We have, for $p=p_c-\varepsilon$ and $\varepsilon\in (0, \, p_c)$,  
 $$
 \liminf_{n\to\infty} \frac1n \log \P(X_n \ge 1)
 \ge 
 - \varepsilon^{\frac12+o(1)},
 $$

\noindent where $o(1)$ goes to $0$ as $\varepsilon\to0$. 

\end{proposition}

\medskip

The rest of the section is devoted to the proof of Proposition \ref{p:lb}. 

While the proof of the upper bound in Theorem \ref{t:main} presented in Section \ref{s:ub} was purely analytic, the proof of its lower bound is probabilistic. It requires a simple hierarchical representation of the system, together with the notion of open paths in the system. 
The main ingredient in the proof of Proposition \ref{p:lb} is a coupling inequality (Theorem \ref{t:coupling} below), connecting $\P(X_n\ge 1)$ (with $p<p_c$) to the Laplace transform of the number of open paths when the system is critical. For the sake of clarity, the coupling inequality and the proof of Proposition \ref{p:lb} are presented in distinct parts.

\subsection{A coupling for subcritical and critical systems}
\label{s:pre}

Recall \eqref{Y_0}: ${\mathscr L}_{Y_0}=(1-p_c)\,\delta_0+p_c \, {\mathscr L}_{X_0^*}$. We are going to construct a coupling for recursive systems $(X_n)$ and $(Y_n)$ in the same probability space such that $X_n \le Y_n$ a.s.\ for all $n\ge 0$, where ${\mathscr L}_{X_0}=(1-p)\,\delta_0 + p \, {\mathscr L}_{X_0^*}$ with $p := p_c -\varepsilon$, $\varepsilon\in (0, \, p_c)$. We use a natural hierarchical representation of Derrida--Retaux systems, as in \cite{collet-eckmann-glaser-martin}, \cite{derrida-retaux}, or \cite{xyz_sustainability}.

Let $\T$ be a (reversed) infinite $m$-ary tree. For any vertex $v$ of $\T$, let $|v|$ denote the generation of $v$ (so $|v|=0$ if the vertex $v$ is in the initial generation). We define a family of random variables $(X(v), Y(v), \, v\in \T)$ as follows. Let $Y(v)$, for $v\in \T$ with $|v|=0$, be i.i.d.\ having the law of $Y_0$.  
Let $Z_0$ be a Bernoulli random variable with $\P(Z_0=1)= 1-\frac\varepsilon{p_c}$ and $\P(Z_0=0)=   \frac\varepsilon{p_c}$. Let $Z(v)$, $|v|=0$, be independent copies of $Z_0$, and independent of $(Y(v), \, |v|=0)$. Define $X(v):= Y(v) Z(v)$ for $|v|=0$, 
so that $X(v)$ is distributed as $X_0$ for $|v|=0$.

For any $v\in \T$ with $|v|\ge 1$, we write $v^{(1)}$, $\ldots$, $v^{(m)}$ for the $m$ parents of $v$ in generation $|v|-1$, and define recursively
\begin{eqnarray}
    X(v)
 &:=& (X(v^{(1)}) + \cdots + X(v^{(m)})-1)^+ \, , 
    \label{Xv}
    \\
    Y(v)
 &:=& (Y(v^{(1)}) + \cdots + Y(v^{(m)})-1)^+ \, . 
    \label{Yv}
\end{eqnarray}

\noindent As such, $X(v)$ and $Y(v)$ are well-defined for all $v\in \T$.

For $n\ge 0$, let $\mathfrak{e}_n$ denote the first lexicographic vertex in the $n$-th generation of $\T$. Let $\T_n$ denote the (reversed) subtree formed by all the ancestors (including $\mathfrak{e}_n$ itself) of $\mathfrak{e}_n$ in the first $n$ generations. See Figure \ref{f:fig1} below for an example. By definition, $X(\mathfrak{e}_n)\le Y(\mathfrak{e}_n)$ a.s.\ for all $n\ge 0$, and $(X(\mathfrak{e}_n), \, n\ge 0)$ (resp.\ $(Y(\mathfrak{e}_n), \, n\ge 0)$) has the same law as $(X_n, \, n\ge 0)$ (resp: $(Y_n, \, n\ge 0)$).     

For $v\in \T$ with $|v|=0$ and integer $n\ge 0$, we write $v_n$ for the unique descendant of $v$ in generation $n$, and call $(v=v_0, \, v_1, \, v_2, \ldots, \, v_n)$ the path in $\T$ from $v$ to $v_n$ (or: leading to $v_n$).\footnote{Degenerate case: when $|v|=0$, the path from $v$ to $v$ is reduced to the singleton $v$.} 


\begin{figure}[h!]
 \begin{center}
\begin{tikzpicture}[line cap=round,line join=round,>=triangle 45,x=1cm,y=1cm, scale=0.45]
\draw [line width=0.8pt] (4,8)-- (4,7); 
\draw [line width=0.8pt] (4,7)-- (5,7);
\draw [line width=0.8pt] (5,7)-- (5,8);
\draw [line width=0.8pt] (5,8)-- (4,8);
\draw [line width=0.8pt] (6,8)-- (6,7);
\draw [line width=0.8pt] (6,7)-- (7,7);
\draw [line width=0.8pt] (7,7)-- (7,8);
\draw [line width=0.8pt] (7,8)-- (6,8);
\draw [line width=0.8pt] (8,8)-- (8,7);
\draw [line width=0.8pt] (8,7)-- (9,7);
\draw [line width=0.8pt] (9,7)-- (9,8);
\draw [line width=0.8pt] (9,8)-- (8,8);
\draw [line width=0.8pt] (10,8)-- (10,7);
\draw [line width=0.8pt] (10,7)-- (11,7);
\draw [line width=0.8pt] (11,7)-- (11,8);
\draw [line width=0.8pt] (11,8)-- (10,8);
\draw [line width=0.8pt] (5,5)-- (5,4);
\draw [line width=0.8pt] (5,4)-- (6,4);
\draw [line width=0.8pt] (6,4)-- (6,5);
\draw [line width=0.8pt] (6,5)-- (5,5);
\draw [line width=0.8pt] (9,5)-- (9,4);
\draw [line width=0.8pt] (9,4)-- (10,4);
\draw [line width=0.8pt] (10,4)-- (10,5);
\draw [line width=0.8pt] (10,5)-- (9,5);
\draw [line width=0.8pt] (7,2)-- (7,1);
\draw [line width=0.8pt] (7,1)-- (8,1);
\draw [line width=0.8pt] (8,1)-- (8,2);
\draw [line width=0.8pt] (8,2)-- (7,2);
\draw [line width=0.8pt] (4.5,7)-- (4.5,6);
\draw [line width=0.8pt] (6.5,7)-- (6.5,6);
\draw [line width=0.8pt] (4.5,6)-- (6.5,6);
\draw [line width=0.8pt] (5.5,6)-- (5.5,5);
\draw [line width=0.8pt] (8.5,7)-- (8.5,6);
\draw [line width=0.8pt] (10.5,7)-- (10.5,6);
\draw [line width=0.8pt] (8.5,6)-- (10.5,6);
\draw [line width=0.8pt] (9.5,6)-- (9.5,5);
\draw [line width=0.8pt] (5.5,4)-- (5.5,3);
\draw [line width=0.8pt] (9.5,4)-- (9.5,3);
\draw [line width=0.8pt] (5.5,3)-- (9.5,3);
\draw (4,8.1) node[anchor=north west] {\scriptsize $1$};
\draw [line width=0.8pt] (4,8)-- (4,7);
\draw [line width=0.8pt] (4,7)-- (5,7);
\draw [line width=0.8pt] (5,7)-- (5,8);
\draw [line width=0.8pt] (5,8)-- (4,8);
\draw [line width=0.8pt] (6,8)-- (6,7);
\draw [line width=0.8pt] (6,7)-- (7,7);
\draw [line width=0.8pt] (7,7)-- (7,8);
\draw [line width=0.8pt] (7,8)-- (6,8);
\draw [line width=0.8pt] (8,8)-- (8,7);
\draw [line width=0.8pt] (8,7)-- (9,7);
\draw [line width=0.8pt] (9,7)-- (9,8);
\draw [line width=0.8pt] (9,8)-- (8,8);
\draw [line width=0.8pt] (10,8)-- (10,7);
\draw [line width=0.8pt] (10,7)-- (11,7);
\draw [line width=0.8pt] (11,7)-- (11,8);
\draw [line width=0.8pt] (11,8)-- (10,8);
\draw [line width=0.8pt] (5,5)-- (5,4);
\draw [line width=0.8pt] (5,4)-- (6,4);
\draw [line width=0.8pt] (6,4)-- (6,5);
\draw [line width=0.8pt] (6,5)-- (5,5);
\draw [line width=0.8pt] (9,5)-- (9,4);
\draw [line width=0.8pt] (9,4)-- (10,4);
\draw [line width=0.8pt] (10,4)-- (10,5);
\draw [line width=0.8pt] (10,5)-- (9,5);
\draw [line width=0.8pt] (7,2)-- (7,1);
\draw [line width=0.8pt] (7,1)-- (8,1);
\draw [line width=0.8pt] (8,1)-- (8,2);
\draw [line width=0.8pt] (8,2)-- (7,2);
\draw [line width=0.8pt] (4.5,7)-- (4.5,6);
\draw [line width=0.8pt] (6.5,7)-- (6.5,6);
\draw [line width=0.8pt] (4.5,6)-- (6.5,6);
\draw [line width=0.8pt] (5.5,6)-- (5.5,5);
\draw [line width=0.8pt] (8.5,7)-- (8.5,6);
\draw [line width=0.8pt] (10.5,7)-- (10.5,6);
\draw [line width=0.8pt] (8.5,6)-- (10.5,6);
\draw [line width=0.8pt] (9.5,6)-- (9.5,5);
\draw [line width=0.8pt] (5.5,4)-- (5.5,3);
\draw [line width=0.8pt] (9.5,4)-- (9.5,3);
\draw [line width=0.8pt] (5.5,3)-- (9.5,3);
\draw [line width=0.8pt] (4,8)-- (4,7);
\draw [line width=0.8pt] (4,7)-- (5,7);
\draw [line width=0.8pt] (5,7)-- (5,8);
\draw [line width=0.8pt] (5,8)-- (4,8);
\draw [line width=0.8pt] (6,8)-- (6,7);
\draw [line width=0.8pt] (6,7)-- (7,7);
\draw [line width=0.8pt] (7,7)-- (7,8);
\draw [line width=0.8pt] (7,8)-- (6,8);
\draw [line width=0.8pt] (8,8)-- (8,7);
\draw [line width=0.8pt] (8,7)-- (9,7);
\draw [line width=0.8pt] (9,7)-- (9,8);
\draw [line width=0.8pt] (9,8)-- (8,8);
\draw [line width=0.8pt] (10,8)-- (10,7);
\draw [line width=0.8pt] (10,7)-- (11,7);
\draw [line width=0.8pt] (11,7)-- (11,8);
\draw [line width=0.8pt] (11,8)-- (10,8);
\draw [line width=0.8pt] (5,5)-- (5,4);
\draw [line width=0.8pt] (5,4)-- (6,4);
\draw [line width=0.8pt] (6,4)-- (6,5);
\draw [line width=0.8pt] (6,5)-- (5,5);
\draw [line width=0.8pt] (9,5)-- (9,4);
\draw [line width=0.8pt] (9,4)-- (10,4);
\draw [line width=0.8pt] (10,4)-- (10,5);
\draw [line width=0.8pt] (10,5)-- (9,5);
\draw [line width=0.8pt] (7,2)-- (7,1);
\draw [line width=0.8pt] (7,1)-- (8,1);
\draw [line width=0.8pt] (8,1)-- (8,2);
\draw [line width=0.8pt] (8,2)-- (7,2);
\draw [line width=0.8pt] (4.5,7)-- (4.5,6);
\draw [line width=0.8pt] (6.5,7)-- (6.5,6);
\draw [line width=0.8pt] (4.5,6)-- (6.5,6);
\draw [line width=0.8pt] (5.5,6)-- (5.5,5);
\draw [line width=0.8pt] (8.5,7)-- (8.5,6);
\draw [line width=0.8pt] (10.5,7)-- (10.5,6);
\draw [line width=0.8pt] (8.5,6)-- (10.5,6);
\draw [line width=0.8pt] (9.5,6)-- (9.5,5);
\draw [line width=0.8pt] (5.5,4)-- (5.5,3);
\draw [line width=0.8pt] (9.5,4)-- (9.5,3);
\draw [line width=0.8pt] (5.5,3)-- (9.5,3);
\draw [line width=0.8pt] (4,8)-- (4,7);
\draw [line width=0.8pt] (4,7)-- (5,7);
\draw [line width=0.8pt] (5,7)-- (5,8);
\draw [line width=0.8pt] (5,8)-- (4,8);
\draw [line width=0.8pt] (6,8)-- (6,7);
\draw [line width=0.8pt] (6,7)-- (7,7);
\draw [line width=0.8pt] (7,7)-- (7,8);
\draw [line width=0.8pt] (7,8)-- (6,8);
\draw [line width=0.8pt] (8,8)-- (8,7);
\draw [line width=0.8pt] (8,7)-- (9,7);
\draw [line width=0.8pt] (9,7)-- (9,8);
\draw [line width=0.8pt] (9,8)-- (8,8);
\draw [line width=0.8pt] (10,8)-- (10,7);
\draw [line width=0.8pt] (10,7)-- (11,7);
\draw [line width=0.8pt] (11,7)-- (11,8);
\draw [line width=0.8pt] (11,8)-- (10,8);
\draw [line width=0.8pt] (5,5)-- (5,4);
\draw [line width=0.8pt] (5,4)-- (6,4);
\draw [line width=0.8pt] (6,4)-- (6,5);
\draw [line width=0.8pt] (6,5)-- (5,5);
\draw [line width=0.8pt] (9,5)-- (9,4);
\draw [line width=0.8pt] (9,4)-- (10,4);
\draw [line width=0.8pt] (10,4)-- (10,5);
\draw [line width=0.8pt] (10,5)-- (9,5);
\draw [line width=0.8pt] (7,2)-- (7,1);
\draw [line width=0.8pt] (7,1)-- (8,1);
\draw [line width=0.8pt] (8,1)-- (8,2);
\draw [line width=0.8pt] (8,2)-- (7,2);
\draw [line width=0.8pt] (4.5,7)-- (4.5,6);
\draw [line width=0.8pt] (6.5,7)-- (6.5,6);
\draw [line width=0.8pt] (4.5,6)-- (6.5,6);
\draw [line width=0.8pt] (5.5,6)-- (5.5,5);
\draw [line width=0.8pt] (8.5,7)-- (8.5,6);
\draw [line width=0.8pt] (10.5,7)-- (10.5,6);
\draw [line width=0.8pt] (8.5,6)-- (10.5,6);
\draw [line width=0.8pt] (9.5,6)-- (9.5,5);
\draw [line width=0.8pt] (5.5,4)-- (5.5,3);
\draw [line width=0.8pt] (9.5,4)-- (9.5,3);
\draw [line width=0.8pt] (5.5,3)-- (9.5,3);
\draw [line width=0.8pt] (4,8)-- (4,7);
\draw [line width=0.8pt] (4,7)-- (5,7);
\draw [line width=0.8pt] (5,7)-- (5,8);
\draw [line width=0.8pt] (5,8)-- (4,8);
\draw [line width=0.8pt] (6,8)-- (6,7);
\draw [line width=0.8pt] (6,7)-- (7,7);
\draw [line width=0.8pt] (7,7)-- (7,8);
\draw [line width=0.8pt] (7,8)-- (6,8);
\draw [line width=0.8pt] (8,8)-- (8,7);
\draw [line width=0.8pt] (8,7)-- (9,7);
\draw [line width=0.8pt] (9,7)-- (9,8);
\draw [line width=0.8pt] (9,8)-- (8,8);
\draw [line width=0.8pt] (10,8)-- (10,7);
\draw [line width=0.8pt] (10,7)-- (11,7);
\draw [line width=0.8pt] (11,7)-- (11,8);
\draw [line width=0.8pt] (11,8)-- (10,8);
\draw [line width=0.8pt] (5,5)-- (5,4);
\draw [line width=0.8pt] (5,4)-- (6,4);
\draw [line width=0.8pt] (6,4)-- (6,5);
\draw [line width=0.8pt] (6,5)-- (5,5);
\draw [line width=0.8pt] (9,5)-- (9,4);
\draw [line width=0.8pt] (9,4)-- (10,4);
\draw [line width=0.8pt] (10,4)-- (10,5);
\draw [line width=0.8pt] (10,5)-- (9,5);
\draw [line width=0.8pt] (13,5)-- (13,4);
\draw [line width=0.8pt] (13,4)-- (14,4);
\draw [line width=0.8pt] (14,4)-- (14,5);
\draw [line width=0.8pt] (14,5)-- (13,5);
\draw [line width=0.8pt] (7,2)-- (7,1);
\draw [line width=0.8pt] (7,1)-- (8,1);
\draw [line width=0.8pt] (8,1)-- (8,2);
\draw [line width=0.8pt] (8,2)-- (7,2);
\draw [line width=0.8pt] (4.5,7)-- (4.5, 6);
\draw [line width=0.8pt] (6.5,7)-- (6.5, 6);
\draw [line width=0.8pt] (4.5,6)-- (6.5,6);
\draw [line width=0.8pt] (5.5,6)-- (5.5,5);
\draw [line width=0.8pt] (8.5,7)-- (8.5,6);
\draw [line width=0.8pt] (10.5,7)-- (10.5,6);
\draw [line width=0.8pt] (8.5,6)-- (10.5,6);
\draw [line width=0.8pt] (9.5,6)-- (9.5,5);
\draw [line width=0.8pt] (5.5,4)-- (5.5,3);
\draw [line width=0.8pt] (9.5,4)-- (9.5,3);
\draw [line width=0.8pt] (5.5,3)-- (9.5,3);
\draw [line width=0.8pt] (13.5,4)-- (13.5,3);
\draw [line width=0.8pt] (20,8)-- (20,7);
\draw [line width=0.8pt] (20,7)-- (21,7);
\draw [line width=0.8pt] (21,7)-- (21,8);
\draw [line width=0.8pt] (21,8)-- (20,8);
\draw [line width=0.8pt] (22,8)-- (22,7);
\draw [line width=0.8pt] (22,7)-- (23,7);
\draw [line width=0.8pt] (23,7)-- (23,8);
\draw [line width=0.8pt] (23,8)-- (22,8);
\draw [line width=0.8pt] (24,8)-- (24,7);
\draw [line width=0.8pt] (24,7)-- (25,7);
\draw [line width=0.8pt] (25,7)-- (25,8);
\draw [line width=0.8pt] (25,8)-- (24,8);
\draw [line width=0.8pt] (26,8)-- (26,7);
\draw [line width=0.8pt] (26,7)-- (27,7);
\draw [line width=0.8pt] (27,7)-- (27,8);
\draw [line width=0.8pt] (27,8)-- (26,8);
\draw [line width=0.8pt] (28,8)-- (28,7);
\draw [line width=0.8pt] (28,7)-- (29,7);
\draw [line width=0.8pt] (29,7)-- (29,8);
\draw [line width=0.8pt] (29,8)-- (28,8);
\draw [line width=0.8pt] (30,8)-- (30,7);
\draw [line width=0.8pt] (30,7)-- (31,7);
\draw [line width=0.8pt] (31,7)-- (31,8);
\draw [line width=0.8pt] (31,8)-- (30,8);
\draw [line width=0.8pt] (32,8)-- (32,7);
\draw [line width=0.8pt] (32,7)-- (33,7);
\draw [line width=0.8pt] (33,7)-- (33,8);
\draw [line width=0.8pt] (33,8)-- (32,8);
\draw [line width=0.8pt] (34,8)-- (34,7);
\draw [line width=0.8pt] (34,7)-- (35,7);
\draw [line width=0.8pt] (35,7)-- (35,8);
\draw [line width=0.8pt] (35,8)-- (34,8);
\draw [line width=0.8pt] (21,5)-- (21,4);
\draw [line width=0.8pt] (21,4)-- (22,4);
\draw [line width=0.8pt] (22,4)-- (22,5);
\draw [line width=0.8pt] (22,5)-- (21,5);
\draw [line width=0.8pt] (25,5)-- (25,4);
\draw [line width=0.8pt] (25,4)-- (26,4);
\draw [line width=0.8pt] (26,4)-- (26,5);
\draw [line width=0.8pt] (26,5)-- (25,5);
\draw [line width=0.8pt] (29,5)-- (29,4);
\draw [line width=0.8pt] (29,4)-- (30,4);
\draw [line width=0.8pt] (30,4)-- (30,5);
\draw [line width=0.8pt] (30,5)-- (29,5);
\draw [line width=0.8pt] (33,5)-- (33,4);
\draw [line width=0.8pt] (33,4)-- (34,4);
\draw [line width=0.8pt] (34,4)-- (34,5);
\draw [line width=0.8pt] (34,5)-- (33,5);
\draw [line width=0.8pt] (23,2)-- (23,1);
\draw [line width=0.8pt] (23,1)-- (24,1);
\draw [line width=0.8pt] (24,1)-- (24,2);
\draw [line width=0.8pt] (24,2)-- (23,2);
\draw [line width=0.8pt] (31,2)-- (31,1);
\draw [line width=0.8pt] (31,1)-- (32,1);
\draw [line width=0.8pt] (32,1)-- (32,2);
\draw [line width=0.8pt] (32,2)-- (31,2);
\draw [line width=0.8pt] (27,-1)-- (27,-2);
\draw [line width=0.8pt] (27,-2)-- (28,-2);
\draw [line width=0.8pt] (28,-2)-- (28,-1);
\draw [line width=0.8pt] (28,-1)-- (27,-1);
\draw [line width=0.8pt] (19,-4)-- (19,-5);
\draw [line width=0.8pt] (19,-5)-- (20,-5);
\draw [line width=0.8pt] (20,-5)-- (20,-4);
\draw [line width=0.8pt] (20,-4)-- (19,-4);
\draw [line width=0.8pt] (11.5,-2)-- (11.5,-3);
\draw [line width=0.8pt] (27.5,-2)-- (27.5,-3);
\draw [line width=0.8pt] (20.5,7)-- (20.5,6);
\draw [line width=0.8pt] (22.5,7)-- (22.5,6);
\draw [line width=0.8pt] (20.5,6)-- (22.5,6);
\draw [line width=0.8pt] (21.5,6)-- (21.5,5);
\draw [line width=0.8pt] (21.5,4)-- (21.5,3);
\draw [line width=0.8pt] (24.5,7)-- (24.5,6);
\draw [line width=0.8pt] (26.5,7)-- (26.5,6);
\draw [line width=0.8pt] (24.5,6)-- (26.5,6);
\draw [line width=0.8pt] (25.5,6)-- (25.5,5);
\draw [line width=0.8pt] (25.5,4)-- (25.5,3);
\draw [line width=0.8pt] (21.5,3)-- (25.5,3);
\draw [line width=0.8pt] (23.5,3)-- (23.5,2);
\draw [line width=0.8pt] (28.5,7)-- (28.5,6);
\draw [line width=0.8pt] (30.5,7)-- (30.5,6);
\draw [line width=0.8pt] (28.5,6)-- (30.5,6);
\draw [line width=0.8pt] (29.5,6)-- (29.5,5);
\draw [line width=0.8pt] (32.5,7)-- (32.5,6);
\draw [line width=0.8pt] (34.5,7)-- (34.5,6);
\draw [line width=0.8pt] (32.5,6)-- (34.5,6);
\draw [line width=0.8pt] (33.5,6)-- (33.5,5);
\draw [line width=0.8pt] (29.5,4)-- (29.5,3);
\draw [line width=0.8pt] (33.5,4)-- (33.5,3);
\draw [line width=0.8pt] (29.5,3)-- (33.5,3);
\draw [line width=0.8pt] (31.5,3)-- (31.5,2);
\draw [line width=0.8pt] (23.5,1)-- (23.5,0);
\draw [line width=0.8pt] (31.5,1)-- (31.5,0);
\draw [line width=0.8pt] (23.5,0)-- (31.5,0);
\draw [line width=0.8pt] (27.5,0)-- (27.5,-1);
\draw (14,8.1) node[anchor=north west] {\scriptsize $2$};
\draw (6.1,8.1) node[anchor=north west] {\scriptsize $0$};
\draw (8,8.1) node[anchor=north west] {\scriptsize $0$};
\draw (10,8.1) node[anchor=north west] {\scriptsize $0$};
\draw (12,8.1) node[anchor=north west] {\scriptsize $3$};
\draw (16,8.1) node[anchor=north west] {\scriptsize $0$};
\draw (18,8.1) node[anchor=north west] {\scriptsize $1$};
\draw (20,8.1) node[anchor=north west] {\scriptsize $0$};
\draw (22,8.1) node[anchor=north west] {\scriptsize $0$};
\draw (24,8.1) node[anchor=north west] {\scriptsize $1$};
\draw (26,8.1) node[anchor=north west] {\scriptsize $0$};
\draw (28,8.1) node[anchor=north west] {\scriptsize $0$};
\draw (30,8.1) node[anchor=north west] {\scriptsize $0$};
\draw (34,8.1) node[anchor=north west] {\scriptsize $0$};
\draw (5,5.1) node[anchor=north west] {\scriptsize $0$};
\draw (9,5.1) node[anchor=north west] {\scriptsize $0$};
\draw (13,5.1) node[anchor=north west] {\scriptsize $4$};
\draw (17,5.1) node[anchor=north west] {\scriptsize $0$};
\draw (7,2.1) node[anchor=north west] {\scriptsize $0$};
\draw (15,2.1) node[anchor=north west] {\scriptsize $3$};
\draw (11,-1) node[anchor=north west] {\scriptsize $2$};
\draw [line width=0.8pt] (11,-1)-- (11,-2);
\draw [line width=0.8pt] (11,-2)-- (12,-2);
\draw [line width=0.8pt] (12,-2)-- (12,-1);
\draw [line width=0.8pt] (12,-1)-- (11,-1);
\draw (32,8.1) node[anchor=north west] {\scriptsize $2$};
\draw (21,5.1) node[anchor=north west] {\scriptsize $0$};
\draw (25,5.1) node[anchor=north west] {\scriptsize $0$};
\draw (23,2.1) node[anchor=north west] {\scriptsize $0$};
\draw (29.05,5.1) node[anchor=north west] {\scriptsize $0$};
\draw (33,5.1) node[anchor=north west] {\scriptsize $1$};
\draw (31,2.1) node[anchor=north west] {\scriptsize $0$};
\draw (27,-1) node[anchor=north west] {\scriptsize $0$};
\draw (19,-4) node[anchor=north west] {\scriptsize $1$};
\draw (2.8,8.2) node[anchor=north west] {${\mathfrak e}_0$};
\draw (3.8,5.1) node[anchor=north west] {${\mathfrak e}_1$};
\draw (5.8,2.1) node[anchor=north west] {${\mathfrak e}_2$};
\draw (9.8,-1) node[anchor=north west] {${\mathfrak e}_3$};
\draw (17.8,-4) node[anchor=north west] {${\mathfrak e}_4$};
\draw (18.8,7.9) node[anchor=north west] {\scriptsize $v_0$};
\draw (18.1,4.9) node[anchor=north west] {\scriptsize $v_1$};
\draw (16.1,1.9) node[anchor=north west] {\scriptsize $v_2$};
\draw (12,-1.2) node[anchor=north west] {\scriptsize $v_3$};
\draw (20,-4.2) node[anchor=north west] {\scriptsize $v_4$};
\draw [line width=0.8pt] (11.5,0)-- (11.5,-1);
\draw [line width=0.8pt] (19.5,-3)-- (19.5,-4);
\draw [line width=0.8pt] (7.5,2)-- (7.5,3);
\draw [line width=0.8pt] (7.5,0)-- (7.5,1);
\draw [line width=0.8pt] (12,7)-- (13,7);
\draw [line width=0.8pt] (13,7)-- (13,8);
\draw [line width=0.8pt] (13,8)-- (12,8);
\draw [line width=0.8pt] (12,8)-- (12,7);
\draw [line width=0.8pt] (12,7)-- (13,7);
\draw [line width=0.8pt] (13,7)-- (13,8);
\draw [line width=0.8pt] (13,8)-- (12,8);
\draw [line width=0.8pt] (12,8)-- (12,7);
\draw [line width=0.8pt] (12,7)-- (13,7);
\draw [line width=0.8pt] (13,7)-- (13,8);
\draw [line width=0.8pt] (13,8)-- (12,8);
\draw [line width=0.8pt] (12,8)-- (12,7);
\draw [line width=0.8pt] (12,7)-- (13,7);
\draw [line width=0.8pt] (13,7)-- (13,8);
\draw [line width=0.8pt] (13,8)-- (12,8);
\draw [line width=0.8pt] (12,8)-- (12,7);
\draw [line width=0.8pt] (12,7)-- (13,7);
\draw [line width=0.8pt] (13,7)-- (13,8);
\draw [line width=0.8pt] (13,8)-- (12,8);
\draw [line width=0.8pt] (12,8)-- (12,7);
\draw [line width=0.8pt] (12,7)-- (13,7);
\draw [line width=0.8pt] (13,7)-- (13,8);
\draw [line width=0.8pt] (13,8)-- (12,8);
\draw [line width=0.8pt] (12,8)-- (12,7);
\draw [line width=0.8pt] (12,7)-- (13,7);
\draw [line width=0.8pt] (13,7)-- (13,8);
\draw [line width=0.8pt] (13,8)-- (12,8);
\draw [line width=0.8pt] (12,8)-- (12,7);
\draw [line width=0.8pt] (12,7)-- (13,7);
\draw [line width=0.8pt] (13,7)-- (13,8);
\draw [line width=0.8pt] (13,8)-- (12,8);
\draw [line width=0.8pt] (12,8)-- (12,7);
\draw [line width=0.8pt] (12,7)-- (13,7);
\draw [line width=0.8pt] (13,7)-- (13,8);
\draw [line width=0.8pt] (13,8)-- (12,8);
\draw [line width=0.8pt] (12,8)-- (12,7);
\draw [line width=0.8pt] (14,7)-- (15,7);
\draw [line width=0.8pt] (15,7)-- (15,8);
\draw [line width=0.8pt] (15,8)-- (14,8);
\draw [line width=0.8pt] (14,8)-- (14,7);
\draw [line width=0.8pt] (16,7)-- (17,7);
\draw [line width=0.8pt] (17,7)-- (17,8);
\draw [line width=0.8pt] (17,8)-- (16,8);
\draw [line width=0.8pt] (16,8)-- (16,7);
\draw [line width=0.8pt] (18,7)-- (19,7);
\draw [line width=0.8pt] (19,7)-- (19,8);
\draw [line width=0.8pt] (19,8)-- (18,8);
\draw [line width=0.8pt] (18,8)-- (18,7);
\draw [line width=0.8pt] (18.5,6)-- (18.5,7);
\draw [line width=0.8pt] (14.5,7)-- (14.5,6);
\draw [line width=0.8pt] (16.5,7)-- (16.5,6);
\draw [line width=0.8pt] (12.5,7)-- (12.5,6);
\draw [line width=0.8pt] (12.5,6)-- (14.5,6);
\draw [line width=0.8pt] (16.5,6)-- (18.5,6);
\draw [line width=0.8pt] (17,4)-- (18,4);
\draw [line width=0.8pt] (18,4)-- (18,5);
\draw [line width=0.8pt] (18,5)-- (17,5);
\draw [line width=0.8pt] (17,5)-- (17,4);
\draw [line width=0.8pt] (13.5,6)-- (13.5,5);
\draw [line width=0.8pt] (17.5,6)-- (17.5,5);
\draw [line width=0.8pt] (17.5,4)-- (17.5,3);
\draw [line width=0.8pt] (13.5,3)-- (17.5,3);
\draw [line width=0.8pt] (15,1)-- (16,1);
\draw [line width=0.8pt] (16,1)-- (16,2);
\draw [line width=0.8pt] (16,2)-- (15,2);
\draw [line width=0.8pt] (15,2)-- (15,1);
\draw [line width=0.8pt] (15.5,3)-- (15.5,2);
\draw [line width=0.8pt] (15.5,1)-- (15.5,0);
\draw [line width=0.8pt] (15.5,0)-- (11.5,0);
\draw [line width=0.8pt] (7.5,0)-- (11.5,0);
\draw [line width=0.8pt] (11.5,-3)-- (19.5,-3);
\draw [line width=0.8pt] (19.5,-3)-- (27.5,-3);
\end{tikzpicture}
\caption{Two paths ${\mathfrak e}_0, ..., {\mathfrak e}_4$ and  $v_0, ..., v_4$ with $v_3={\mathfrak e}_3$ and $v_4={\mathfrak e}_4$.}
  \label{f:fig1}
\end{center}
\end{figure}

For any $u \in \T$, let $u_*$ be the unique child of $u$. Denote by $\mathtt{bro}(u)$ the set of the ``brothers" of $u$, i.e., the parents of $u_*$ that are not $u$. Let
$$
\xi(u)
:=
\sum_{y\in \mathtt{bro}(u)} Y(y).
$$ 

\noindent The path $(v=v_0, \, v_1, \, v_2, \ldots, \, v_n)$ is called {\it open} (for the critical system $(Y_n)$) if 
$$
Y(v) +\xi(v_0) + \xi(v_1) + \cdots + \xi(v_i)  \ge i+1, 
\qquad
\forall\,  0\le i\le n-1.
$$

\noindent Define for any $u$ in the tree $\T$, 
\begin{equation}
    N(u)
    := 
    \mbox{number of open paths from the initial generation to $u$}. 
    \label{def-N} 
\end{equation}

\noindent In other words, if $n:= |u|$, then
$$
N(u) := \sum_{v\in \T: \, |v|=0, \, v_n=u} {\bf 1}_{\{ \hbox{\scriptsize $(v=v_0, \, v_1, \, v_2, \ldots, \, v_n=u)$ is open}\}} \, .
$$

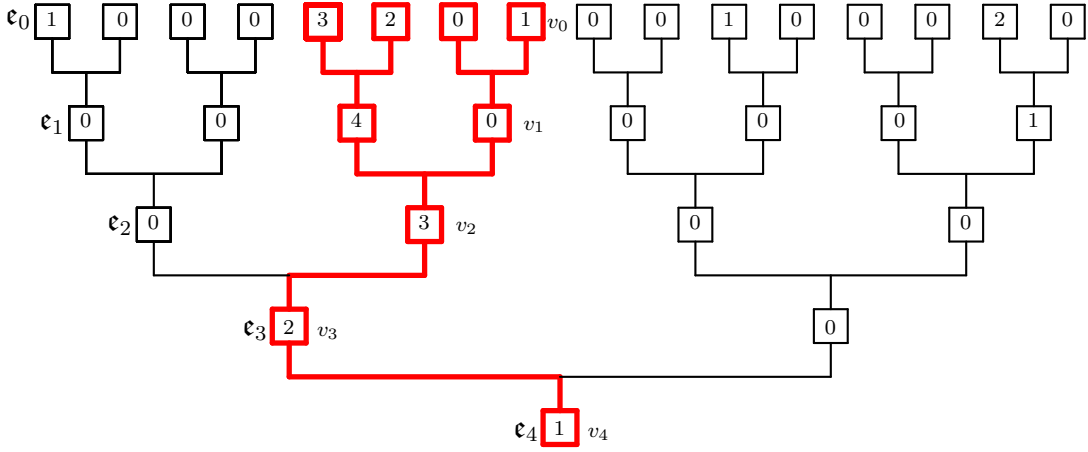
\begin{figure}[h!]
 \begin{center}
\definecolor{ffqqqq}{rgb}{1,0,0}
\begin{tikzpicture}[line cap=round,line join=round,>=triangle 45,x=1cm,y=1cm, scale=0.45]
\draw [line width=0.8pt] (4,8)-- (4,7); 
\draw [line width=0.8pt] (4,7)-- (5,7);
\draw [line width=0.8pt] (5,7)-- (5,8);
\draw [line width=0.8pt] (5,8)-- (4,8);
\draw [line width=0.8pt] (6,8)-- (6,7);
\draw [line width=0.8pt] (6,7)-- (7,7);
\draw [line width=0.8pt] (7,7)-- (7,8);
\draw [line width=0.8pt] (7,8)-- (6,8);
\draw [line width=0.8pt] (8,8)-- (8,7);
\draw [line width=0.8pt] (8,7)-- (9,7);
\draw [line width=0.8pt] (9,7)-- (9,8);
\draw [line width=0.8pt] (9,8)-- (8,8);
\draw [line width=0.8pt] (10,8)-- (10,7);
\draw [line width=0.8pt] (10,7)-- (11,7);
\draw [line width=0.8pt] (11,7)-- (11,8);
\draw [line width=0.8pt] (11,8)-- (10,8);
\draw [line width=0.8pt] (5,5)-- (5,4);
\draw [line width=0.8pt] (5,4)-- (6,4);
\draw [line width=0.8pt] (6,4)-- (6,5);
\draw [line width=0.8pt] (6,5)-- (5,5);
\draw [line width=0.8pt] (9,5)-- (9,4);
\draw [line width=0.8pt] (9,4)-- (10,4);
\draw [line width=0.8pt] (10,4)-- (10,5);
\draw [line width=0.8pt] (10,5)-- (9,5);
\draw [line width=0.8pt] (7,2)-- (7,1);
\draw [line width=0.8pt] (7,1)-- (8,1);
\draw [line width=0.8pt] (8,1)-- (8,2);
\draw [line width=0.8pt] (8,2)-- (7,2);
\draw [line width=0.8pt] (4.5,7)-- (4.5,6);
\draw [line width=0.8pt] (6.5,7)-- (6.5,6);
\draw [line width=0.8pt] (4.5,6)-- (6.5,6);
\draw [line width=0.8pt] (5.5,6)-- (5.5,5);
\draw [line width=0.8pt] (8.5,7)-- (8.5,6);
\draw [line width=0.8pt] (10.5,7)-- (10.5,6);
\draw [line width=0.8pt] (8.5,6)-- (10.5,6);
\draw [line width=0.8pt] (9.5,6)-- (9.5,5);
\draw [line width=0.8pt] (5.5,4)-- (5.5,3);
\draw [line width=0.8pt] (9.5,4)-- (9.5,3);
\draw [line width=0.8pt] (5.5,3)-- (9.5,3);
\draw (4,8.1) node[anchor=north west] {\scriptsize $1$};
\draw [line width=0.8pt] (4,8)-- (4,7);
\draw [line width=0.8pt] (4,7)-- (5,7);
\draw [line width=0.8pt] (5,7)-- (5,8);
\draw [line width=0.8pt] (5,8)-- (4,8);
\draw [line width=0.8pt] (6,8)-- (6,7);
\draw [line width=0.8pt] (6,7)-- (7,7);
\draw [line width=0.8pt] (7,7)-- (7,8);
\draw [line width=0.8pt] (7,8)-- (6,8);
\draw [line width=0.8pt] (8,8)-- (8,7);
\draw [line width=0.8pt] (8,7)-- (9,7);
\draw [line width=0.8pt] (9,7)-- (9,8);
\draw [line width=0.8pt] (9,8)-- (8,8);
\draw [line width=0.8pt] (10,8)-- (10,7);
\draw [line width=0.8pt] (10,7)-- (11,7);
\draw [line width=0.8pt] (11,7)-- (11,8);
\draw [line width=0.8pt] (11,8)-- (10,8);
\draw [line width=0.8pt] (5,5)-- (5,4);
\draw [line width=0.8pt] (5,4)-- (6,4);
\draw [line width=0.8pt] (6,4)-- (6,5);
\draw [line width=0.8pt] (6,5)-- (5,5);
\draw [line width=0.8pt] (9,5)-- (9,4);
\draw [line width=0.8pt] (9,4)-- (10,4);
\draw [line width=0.8pt] (10,4)-- (10,5);
\draw [line width=0.8pt] (10,5)-- (9,5);
\draw [line width=0.8pt] (7,2)-- (7,1);
\draw [line width=0.8pt] (7,1)-- (8,1);
\draw [line width=0.8pt] (8,1)-- (8,2);
\draw [line width=0.8pt] (8,2)-- (7,2);
\draw [line width=0.8pt] (4.5,7)-- (4.5,6);
\draw [line width=0.8pt] (6.5,7)-- (6.5,6);
\draw [line width=0.8pt] (4.5,6)-- (6.5,6);
\draw [line width=0.8pt] (5.5,6)-- (5.5,5);
\draw [line width=0.8pt] (8.5,7)-- (8.5,6);
\draw [line width=0.8pt] (10.5,7)-- (10.5,6);
\draw [line width=0.8pt] (8.5,6)-- (10.5,6);
\draw [line width=0.8pt] (9.5,6)-- (9.5,5);
\draw [line width=0.8pt] (5.5,4)-- (5.5,3);
\draw [line width=0.8pt] (9.5,4)-- (9.5,3);
\draw [line width=0.8pt] (5.5,3)-- (9.5,3);
\draw [line width=0.8pt] (4,8)-- (4,7);
\draw [line width=0.8pt] (4,7)-- (5,7);
\draw [line width=0.8pt] (5,7)-- (5,8);
\draw [line width=0.8pt] (5,8)-- (4,8);
\draw [line width=0.8pt] (6,8)-- (6,7);
\draw [line width=0.8pt] (6,7)-- (7,7);
\draw [line width=0.8pt] (7,7)-- (7,8);
\draw [line width=0.8pt] (7,8)-- (6,8);
\draw [line width=0.8pt] (8,8)-- (8,7);
\draw [line width=0.8pt] (8,7)-- (9,7);
\draw [line width=0.8pt] (9,7)-- (9,8);
\draw [line width=0.8pt] (9,8)-- (8,8);
\draw [line width=0.8pt] (10,8)-- (10,7);
\draw [line width=0.8pt] (10,7)-- (11,7);
\draw [line width=0.8pt] (11,7)-- (11,8);
\draw [line width=0.8pt] (11,8)-- (10,8);
\draw [line width=0.8pt] (5,5)-- (5,4);
\draw [line width=0.8pt] (5,4)-- (6,4);
\draw [line width=0.8pt] (6,4)-- (6,5);
\draw [line width=0.8pt] (6,5)-- (5,5);
\draw [line width=0.8pt] (9,5)-- (9,4);
\draw [line width=0.8pt] (9,4)-- (10,4);
\draw [line width=0.8pt] (10,4)-- (10,5);
\draw [line width=0.8pt] (10,5)-- (9,5);
\draw [line width=0.8pt] (7,2)-- (7,1);
\draw [line width=0.8pt] (7,1)-- (8,1);
\draw [line width=0.8pt] (8,1)-- (8,2);
\draw [line width=0.8pt] (8,2)-- (7,2);
\draw [line width=0.8pt] (4.5,7)-- (4.5,6);
\draw [line width=0.8pt] (6.5,7)-- (6.5,6);
\draw [line width=0.8pt] (4.5,6)-- (6.5,6);
\draw [line width=0.8pt] (5.5,6)-- (5.5,5);
\draw [line width=0.8pt] (8.5,7)-- (8.5,6);
\draw [line width=0.8pt] (10.5,7)-- (10.5,6);
\draw [line width=0.8pt] (8.5,6)-- (10.5,6);
\draw [line width=0.8pt] (9.5,6)-- (9.5,5);
\draw [line width=0.8pt] (5.5,4)-- (5.5,3);
\draw [line width=0.8pt] (9.5,4)-- (9.5,3);
\draw [line width=0.8pt] (5.5,3)-- (9.5,3);
\draw [line width=0.8pt] (4,8)-- (4,7);
\draw [line width=0.8pt] (4,7)-- (5,7);
\draw [line width=0.8pt] (5,7)-- (5,8);
\draw [line width=0.8pt] (5,8)-- (4,8);
\draw [line width=0.8pt] (6,8)-- (6,7);
\draw [line width=0.8pt] (6,7)-- (7,7);
\draw [line width=0.8pt] (7,7)-- (7,8);
\draw [line width=0.8pt] (7,8)-- (6,8);
\draw [line width=0.8pt] (8,8)-- (8,7);
\draw [line width=0.8pt] (8,7)-- (9,7);
\draw [line width=0.8pt] (9,7)-- (9,8);
\draw [line width=0.8pt] (9,8)-- (8,8);
\draw [line width=0.8pt] (10,8)-- (10,7);
\draw [line width=0.8pt] (10,7)-- (11,7);
\draw [line width=0.8pt] (11,7)-- (11,8);
\draw [line width=0.8pt] (11,8)-- (10,8);
\draw [line width=0.8pt] (5,5)-- (5,4);
\draw [line width=0.8pt] (5,4)-- (6,4);
\draw [line width=0.8pt] (6,4)-- (6,5);
\draw [line width=0.8pt] (6,5)-- (5,5);
\draw [line width=0.8pt] (9,5)-- (9,4);
\draw [line width=0.8pt] (9,4)-- (10,4);
\draw [line width=0.8pt] (10,4)-- (10,5);
\draw [line width=0.8pt] (10,5)-- (9,5);
\draw [line width=0.8pt] (7,2)-- (7,1);
\draw [line width=0.8pt] (7,1)-- (8,1);
\draw [line width=0.8pt] (8,1)-- (8,2);
\draw [line width=0.8pt] (8,2)-- (7,2);
\draw [line width=0.8pt] (4.5,7)-- (4.5,6);
\draw [line width=0.8pt] (6.5,7)-- (6.5,6);
\draw [line width=0.8pt] (4.5,6)-- (6.5,6);
\draw [line width=0.8pt] (5.5,6)-- (5.5,5);
\draw [line width=0.8pt] (8.5,7)-- (8.5,6);
\draw [line width=0.8pt] (10.5,7)-- (10.5,6);
\draw [line width=0.8pt] (8.5,6)-- (10.5,6);
\draw [line width=0.8pt] (9.5,6)-- (9.5,5);
\draw [line width=0.8pt] (5.5,4)-- (5.5,3);
\draw [line width=0.8pt] (9.5,4)-- (9.5,3);
\draw [line width=0.8pt] (5.5,3)-- (9.5,3);
\draw [line width=0.8pt] (4,8)-- (4,7);
\draw [line width=0.8pt] (4,7)-- (5,7);
\draw [line width=0.8pt] (5,7)-- (5,8);
\draw [line width=0.8pt] (5,8)-- (4,8);
\draw [line width=0.8pt] (6,8)-- (6,7);
\draw [line width=0.8pt] (6,7)-- (7,7);
\draw [line width=0.8pt] (7,7)-- (7,8);
\draw [line width=0.8pt] (7,8)-- (6,8);
\draw [line width=0.8pt] (8,8)-- (8,7);
\draw [line width=0.8pt] (8,7)-- (9,7);
\draw [line width=0.8pt] (9,7)-- (9,8);
\draw [line width=0.8pt] (9,8)-- (8,8);
\draw [line width=0.8pt] (10,8)-- (10,7);
\draw [line width=0.8pt] (10,7)-- (11,7);
\draw [line width=0.8pt] (11,7)-- (11,8);
\draw [line width=0.8pt] (11,8)-- (10,8);
\draw [line width=0.8pt] (5,5)-- (5,4);
\draw [line width=0.8pt] (5,4)-- (6,4);
\draw [line width=0.8pt] (6,4)-- (6,5);
\draw [line width=0.8pt] (6,5)-- (5,5);
\draw [line width=0.8pt] (9,5)-- (9,4);
\draw [line width=0.8pt] (9,4)-- (10,4);
\draw [line width=0.8pt] (10,4)-- (10,5);
\draw [line width=0.8pt] (10,5)-- (9,5);
\draw [line width=2pt,color=ffqqqq] (13,5)-- (13,4);
\draw [line width=2pt,color=ffqqqq] (13,4)-- (14,4);
\draw [line width=2pt,color=ffqqqq] (14,4)-- (14,5);
\draw [line width=2pt,color=ffqqqq] (14,5)-- (13,5);
\draw [line width=0.8pt] (7,2)-- (7,1);
\draw [line width=0.8pt] (7,1)-- (8,1);
\draw [line width=0.8pt] (8,1)-- (8,2);
\draw [line width=0.8pt] (8,2)-- (7,2);
\draw [line width=0.8pt] (4.5,7)-- (4.5, 6);
\draw [line width=0.8pt] (6.5,7)-- (6.5, 6);
\draw [line width=0.8pt] (4.5,6)-- (6.5,6);
\draw [line width=0.8pt] (5.5,6)-- (5.5,5);
\draw [line width=0.8pt] (8.5,7)-- (8.5,6);
\draw [line width=0.8pt] (10.5,7)-- (10.5,6);
\draw [line width=0.8pt] (8.5,6)-- (10.5,6);
\draw [line width=0.8pt] (9.5,6)-- (9.5,5);
\draw [line width=0.8pt] (5.5,4)-- (5.5,3);
\draw [line width=0.8pt] (9.5,4)-- (9.5,3);
\draw [line width=0.8pt] (5.5,3)-- (9.5,3);
\draw [line width=2pt,color=ffqqqq] (13.5,4)-- (13.5,3);
\draw [line width=0.8pt] (20,8)-- (20,7);
\draw [line width=0.8pt] (20,7)-- (21,7);
\draw [line width=0.8pt] (21,7)-- (21,8);
\draw [line width=0.8pt] (21,8)-- (20,8);
\draw [line width=0.8pt] (22,8)-- (22,7);
\draw [line width=0.8pt] (22,7)-- (23,7);
\draw [line width=0.8pt] (23,7)-- (23,8);
\draw [line width=0.8pt] (23,8)-- (22,8);
\draw [line width=0.8pt] (24,8)-- (24,7);
\draw [line width=0.8pt] (24,7)-- (25,7);
\draw [line width=0.8pt] (25,7)-- (25,8);
\draw [line width=0.8pt] (25,8)-- (24,8);
\draw [line width=0.8pt] (26,8)-- (26,7);
\draw [line width=0.8pt] (26,7)-- (27,7);
\draw [line width=0.8pt] (27,7)-- (27,8);
\draw [line width=0.8pt] (27,8)-- (26,8);
\draw [line width=0.8pt] (28,8)-- (28,7);
\draw [line width=0.8pt] (28,7)-- (29,7);
\draw [line width=0.8pt] (29,7)-- (29,8);
\draw [line width=0.8pt] (29,8)-- (28,8);
\draw [line width=0.8pt] (30,8)-- (30,7);
\draw [line width=0.8pt] (30,7)-- (31,7);
\draw [line width=0.8pt] (31,7)-- (31,8);
\draw [line width=0.8pt] (31,8)-- (30,8);
\draw [line width=0.8pt] (32,8)-- (32,7);
\draw [line width=0.8pt] (32,7)-- (33,7);
\draw [line width=0.8pt] (33,7)-- (33,8);
\draw [line width=0.8pt] (33,8)-- (32,8);
\draw [line width=0.8pt] (34,8)-- (34,7);
\draw [line width=0.8pt] (34,7)-- (35,7);
\draw [line width=0.8pt] (35,7)-- (35,8);
\draw [line width=0.8pt] (35,8)-- (34,8);
\draw [line width=0.8pt] (21,5)-- (21,4);
\draw [line width=0.8pt] (21,4)-- (22,4);
\draw [line width=0.8pt] (22,4)-- (22,5);
\draw [line width=0.8pt] (22,5)-- (21,5);
\draw [line width=0.8pt] (25,5)-- (25,4);
\draw [line width=0.8pt] (25,4)-- (26,4);
\draw [line width=0.8pt] (26,4)-- (26,5);
\draw [line width=0.8pt] (26,5)-- (25,5);
\draw [line width=0.8pt] (29,5)-- (29,4);
\draw [line width=0.8pt] (29,4)-- (30,4);
\draw [line width=0.8pt] (30,4)-- (30,5);
\draw [line width=0.8pt] (30,5)-- (29,5);
\draw [line width=0.8pt] (33,5)-- (33,4);
\draw [line width=0.8pt] (33,4)-- (34,4);
\draw [line width=0.8pt] (34,4)-- (34,5);
\draw [line width=0.8pt] (34,5)-- (33,5);
\draw [line width=0.8pt] (23,2)-- (23,1);
\draw [line width=0.8pt] (23,1)-- (24,1);
\draw [line width=0.8pt] (24,1)-- (24,2);
\draw [line width=0.8pt] (24,2)-- (23,2);
\draw [line width=0.8pt] (31,2)-- (31,1);
\draw [line width=0.8pt] (31,1)-- (32,1);
\draw [line width=0.8pt] (32,1)-- (32,2);
\draw [line width=0.8pt] (32,2)-- (31,2);
\draw [line width=0.8pt] (27,-1)-- (27,-2);
\draw [line width=0.8pt] (27,-2)-- (28,-2);
\draw [line width=0.8pt] (28,-2)-- (28,-1);
\draw [line width=0.8pt] (28,-1)-- (27,-1);
\draw [line width=2pt,color=ffqqqq] (19,-4)-- (19,-5);
\draw [line width=2pt,color=ffqqqq] (19,-5)-- (20,-5);
\draw [line width=2pt,color=ffqqqq] (20,-5)-- (20,-4);
\draw [line width=2pt,color=ffqqqq] (20,-4)-- (19,-4);
\draw [line width=2pt,color=ffqqqq] (11.5,-2)-- (11.5,-3);
\draw [line width=0.8pt] (27.5,-2)-- (27.5,-3);
\draw [line width=0.8pt] (20.5,7)-- (20.5,6);
\draw [line width=0.8pt] (22.5,7)-- (22.5,6);
\draw [line width=0.8pt] (20.5,6)-- (22.5,6);
\draw [line width=0.8pt] (21.5,6)-- (21.5,5);
\draw [line width=0.8pt] (21.5,4)-- (21.5,3);
\draw [line width=0.8pt] (24.5,7)-- (24.5,6);
\draw [line width=0.8pt] (26.5,7)-- (26.5,6);
\draw [line width=0.8pt] (24.5,6)-- (26.5,6);
\draw [line width=0.8pt] (25.5,6)-- (25.5,5);
\draw [line width=0.8pt] (25.5,4)-- (25.5,3);
\draw [line width=0.8pt] (21.5,3)-- (25.5,3);
\draw [line width=0.8pt] (23.5,3)-- (23.5,2);
\draw [line width=0.8pt] (28.5,7)-- (28.5,6);
\draw [line width=0.8pt] (30.5,7)-- (30.5,6);
\draw [line width=0.8pt] (28.5,6)-- (30.5,6);
\draw [line width=0.8pt] (29.5,6)-- (29.5,5);
\draw [line width=0.8pt] (32.5,7)-- (32.5,6);
\draw [line width=0.8pt] (34.5,7)-- (34.5,6);
\draw [line width=0.8pt] (32.5,6)-- (34.5,6);
\draw [line width=0.8pt] (33.5,6)-- (33.5,5);
\draw [line width=0.8pt] (29.5,4)-- (29.5,3);
\draw [line width=0.8pt] (33.5,4)-- (33.5,3);
\draw [line width=0.8pt] (29.5,3)-- (33.5,3);
\draw [line width=0.8pt] (31.5,3)-- (31.5,2);
\draw [line width=0.8pt] (23.5,1)-- (23.5,0);
\draw [line width=0.8pt] (31.5,1)-- (31.5,0);
\draw [line width=0.8pt] (23.5,0)-- (31.5,0);
\draw [line width=0.8pt] (27.5,0)-- (27.5,-1);
\draw (14,8.1) node[anchor=north west] {\scriptsize $2$};
\draw (6.1,8.1) node[anchor=north west] {\scriptsize $0$};
\draw (8,8.1) node[anchor=north west] {\scriptsize $0$};
\draw (10,8.1) node[anchor=north west] {\scriptsize $0$};
\draw (12,8.1) node[anchor=north west] {\scriptsize $3$};
\draw (16,8.1) node[anchor=north west] {\scriptsize $0$};
\draw (18,8.1) node[anchor=north west] {\scriptsize $1$};
\draw (20,8.1) node[anchor=north west] {\scriptsize $0$};
\draw (22,8.1) node[anchor=north west] {\scriptsize $0$};
\draw (24,8.1) node[anchor=north west] {\scriptsize $1$};
\draw (26,8.1) node[anchor=north west] {\scriptsize $0$};
\draw (28,8.1) node[anchor=north west] {\scriptsize $0$};
\draw (30,8.1) node[anchor=north west] {\scriptsize $0$};
\draw (34,8.1) node[anchor=north west] {\scriptsize $0$};
\draw (5,5.1) node[anchor=north west] {\scriptsize $0$};
\draw (9,5.1) node[anchor=north west] {\scriptsize $0$};
\draw (13,5.1) node[anchor=north west] {\scriptsize $4$};
\draw (17,5.1) node[anchor=north west] {\scriptsize $0$};
\draw (7,2.1) node[anchor=north west] {\scriptsize $0$};
\draw (15,2.1) node[anchor=north west] {\scriptsize $3$};
\draw (11,-1) node[anchor=north west] {\scriptsize $2$};
\draw [line width=2pt,color=ffqqqq] (11,-1)-- (11,-2);
\draw [line width=2pt,color=ffqqqq] (11,-2)-- (12,-2);
\draw [line width=2pt,color=ffqqqq] (12,-2)-- (12,-1);
\draw [line width=2pt,color=ffqqqq] (12,-1)-- (11,-1);
\draw (32,8.1) node[anchor=north west] {\scriptsize $2$};
\draw (21,5.1) node[anchor=north west] {\scriptsize $0$};
\draw (25,5.1) node[anchor=north west] {\scriptsize $0$};
\draw (23,2.1) node[anchor=north west] {\scriptsize $0$};
\draw (29.05,5.1) node[anchor=north west] {\scriptsize $0$};
\draw (33,5.1) node[anchor=north west] {\scriptsize $1$};
\draw (31,2.1) node[anchor=north west] {\scriptsize $0$};
\draw (27,-1) node[anchor=north west] {\scriptsize $0$};
\draw (19,-4) node[anchor=north west] {\scriptsize $1$};
\draw (2.8,8.2) node[anchor=north west] {${\mathfrak e}_0$};
\draw (3.8,5.1) node[anchor=north west] {${\mathfrak e}_1$};
\draw (5.8,2.1) node[anchor=north west] {${\mathfrak e}_2$};
\draw (9.8,-1) node[anchor=north west] {${\mathfrak e}_3$};
\draw (17.8,-4) node[anchor=north west] {${\mathfrak e}_4$};
\draw (18.8,7.9) node[anchor=north west] {\scriptsize $v_0$};
\draw (18.1,4.9) node[anchor=north west] {\scriptsize $v_1$};
\draw (16.1,1.9) node[anchor=north west] {\scriptsize $v_2$};
\draw (12,-1.2) node[anchor=north west] {\scriptsize $v_3$};
\draw (20,-4.2) node[anchor=north west] {\scriptsize $v_4$};
\draw [line width=2pt,color=ffqqqq] (11.5,0)-- (11.5,-1);
\draw [line width=2pt,color=ffqqqq] (19.5,-3)-- (19.5,-4);
\draw [line width=0.8pt] (7.5,2)-- (7.5,3);
\draw [line width=0.8pt] (7.5,0)-- (7.5,1);
\draw [line width=2pt,color=ffqqqq] (12,7)-- (13,7);
\draw [line width=2pt,color=ffqqqq] (13,7)-- (13,8);
\draw [line width=2pt,color=ffqqqq] (13,8)-- (12,8);
\draw [line width=2pt,color=ffqqqq] (12,8)-- (12,7);
\draw [line width=2pt,color=ffqqqq] (12,7)-- (13,7);
\draw [line width=2pt,color=ffqqqq] (13,7)-- (13,8);
\draw [line width=2pt,color=ffqqqq] (13,8)-- (12,8);
\draw [line width=2pt,color=ffqqqq] (12,8)-- (12,7);
\draw [line width=2pt,color=ffqqqq] (12,7)-- (13,7);
\draw [line width=2pt,color=ffqqqq] (13,7)-- (13,8);
\draw [line width=2pt,color=ffqqqq] (13,8)-- (12,8);
\draw [line width=2pt,color=ffqqqq] (12,8)-- (12,7);
\draw [line width=2pt,color=ffqqqq] (12,7)-- (13,7);
\draw [line width=2pt,color=ffqqqq] (13,7)-- (13,8);
\draw [line width=2pt,color=ffqqqq] (13,8)-- (12,8);
\draw [line width=2pt,color=ffqqqq] (12,8)-- (12,7);
\draw [line width=2pt,color=ffqqqq] (12,7)-- (13,7);
\draw [line width=2pt,color=ffqqqq] (13,7)-- (13,8);
\draw [line width=2pt,color=ffqqqq] (13,8)-- (12,8);
\draw [line width=2pt,color=ffqqqq] (12,8)-- (12,7);
\draw [line width=2pt,color=ffqqqq] (12,7)-- (13,7);
\draw [line width=2pt,color=ffqqqq] (13,7)-- (13,8);
\draw [line width=2pt,color=ffqqqq] (13,8)-- (12,8);
\draw [line width=2pt,color=ffqqqq] (12,8)-- (12,7);
\draw [line width=2pt,color=ffqqqq] (12,7)-- (13,7);
\draw [line width=2pt,color=ffqqqq] (13,7)-- (13,8);
\draw [line width=2pt,color=ffqqqq] (13,8)-- (12,8);
\draw [line width=2pt,color=ffqqqq] (12,8)-- (12,7);
\draw [line width=2pt,color=ffqqqq] (12,7)-- (13,7);
\draw [line width=2pt,color=ffqqqq] (13,7)-- (13,8);
\draw [line width=2pt,color=ffqqqq] (13,8)-- (12,8);
\draw [line width=2pt,color=ffqqqq] (12,8)-- (12,7);
\draw [line width=2pt,color=ffqqqq] (12,7)-- (13,7);
\draw [line width=2pt,color=ffqqqq] (13,7)-- (13,8);
\draw [line width=2pt,color=ffqqqq] (13,8)-- (12,8);
\draw [line width=2pt,color=ffqqqq] (12,8)-- (12,7);
\draw [line width=2pt,color=ffqqqq] (14,7)-- (15,7);
\draw [line width=2pt,color=ffqqqq] (15,7)-- (15,8);
\draw [line width=2pt,color=ffqqqq] (15,8)-- (14,8);
\draw [line width=2pt,color=ffqqqq] (14,8)-- (14,7);
\draw [line width=2pt,color=ffqqqq] (16,7)-- (17,7);
\draw [line width=2pt,color=ffqqqq] (17,7)-- (17,8);
\draw [line width=2pt,color=ffqqqq] (17,8)-- (16,8);
\draw [line width=2pt,color=ffqqqq] (16,8)-- (16,7);
\draw [line width=2pt,color=ffqqqq] (18,7)-- (19,7);
\draw [line width=2pt,color=ffqqqq] (19,7)-- (19,8);
\draw [line width=2pt,color=ffqqqq] (19,8)-- (18,8);
\draw [line width=2pt,color=ffqqqq] (18,8)-- (18,7);
\draw [line width=2pt,color=ffqqqq] (18.5,6)-- (18.5,7);
\draw [line width=2pt,color=ffqqqq] (14.5,7)-- (14.5,6);
\draw [line width=2pt,color=ffqqqq] (16.5,7)-- (16.5,6);
\draw [line width=2pt,color=ffqqqq] (12.5,7)-- (12.5,6);
\draw [line width=2pt,color=ffqqqq] (12.5,6)-- (14.5,6);
\draw [line width=2pt,color=ffqqqq] (16.5,6)-- (18.5,6);
\draw [line width=2pt,color=ffqqqq] (17,4)-- (18,4);
\draw [line width=2pt,color=ffqqqq] (18,4)-- (18,5);
\draw [line width=2pt,color=ffqqqq] (18,5)-- (17,5);
\draw [line width=2pt,color=ffqqqq] (17,5)-- (17,4);
\draw [line width=2pt,color=ffqqqq] (13.5,6)-- (13.5,5);
\draw [line width=2pt,color=ffqqqq] (17.5,6)-- (17.5,5);
\draw [line width=2pt,color=ffqqqq] (17.5,4)-- (17.5,3);
\draw [line width=2pt,color=ffqqqq] (13.5,3)-- (17.5,3);
\draw [line width=2pt,color=ffqqqq] (15,1)-- (16,1);
\draw [line width=2pt,color=ffqqqq] (16,1)-- (16,2);
\draw [line width=2pt,color=ffqqqq] (16,2)-- (15,2);
\draw [line width=2pt,color=ffqqqq] (15,2)-- (15,1);
\draw [line width=2pt,color=ffqqqq] (15.5,3)-- (15.5,2);
\draw [line width=2pt,color=ffqqqq] (15.5,1)-- (15.5,0);
\draw [line width=2pt,color=ffqqqq] (15.5,0)-- (11.5,0);
\draw [line width=0.8pt] (7.5,0)-- (11.5,0);
\draw [line width=2pt,color=ffqqqq] (11.5,-3)-- (19.5,-3);
\draw [line width=0.8pt] (19.5,-3)-- (27.5,-3);
\end{tikzpicture}
\end{center}
\caption{\leftskip=1.6truecm \rightskip=1.6truecm   Open paths from the initial generation to $\mathfrak{e}_4$ are marked in bold (and coloured in red), with $N_4=4$ and $Y_4=1$.}
  \label{f:fig2}
\end{figure}

Let $N_n:= N({\mathfrak e}_n)$ and $Y_n:= Y({\mathfrak e}_n)$, $X_n=X(\mathfrak{e}_n)$ for any $n\ge 0$.\footnote{The number of open paths can obviously be defined for any system. In this paper, however, we make use of the number of open paths only for critical systems; this is the reason for which it is defined only for $(Y_n)$.} See Figure \ref{f:fig2} for an example.


The following result connects 
$\P(X_n\ge 1)$ for subcritical systems to the Laplace transform of $N_n$ for critical systems $(Y_n)$. Recall that $\E(X_0^*\,  m^{X_0^*})<\infty$ means $p_c >0$ (see Theorem A in the introduction).

\medskip

\begin{theorem}
 \label{t:coupling}

 Assume $\E(X_0^*\, m^{X_0^*})<\infty$. We have, for $p\in (0, \, p_c)$ and $n\ge 1$,
 \begin{equation} 
     \P(X_n \ge 1)
     \ge 
     \E\Big[ \Big(\frac{p}{p_c} \Big)^{\! N_n} \, {\bf 1}_{\{ Y_n \ge 1\}}\Big] \, .
     \label{XnYnNn}
 \end{equation}

\end{theorem}

\medskip

\noindent {\it Proof of Theorem \ref{t:coupling}.} 
Let ${\mathscr N}_n$ be the set of vertices $v$ with $|v|=0$ such that $v_n = \mathfrak{e}_n$ and that the path $(v=v_0, \, v_1, \, v_2, \ldots, \, v_n=\mathfrak{e}_n)$ is open. So $N_n$ is exactly the cardinality of ${\mathscr N}_n$. 

Since $(Z(v), \, |v|=0)$ and $(Y(v), \, |v|=0)$ are independent, and $\P(Z(v)=1)= \frac{p}{p_c}$ for any $|v|=0$, it follows that 
$$
\P \Big( \bigcap_{v\in {\mathscr N}_n} \{ Z(v)=1\} \, \Big| \, {\mathscr N}_n, \, Y_n \Big) 
=
\Big( \frac{p}{p_c} \Big)^{\! N_n} \, .
$$

\noindent By definition, if $Z(v)=1$ for all $v\in {\mathscr N}_n$, then $X_n = Y_n$. Consequently, 
$$
\P(X_n \ge 1)
\ge
\P \Big( Y_n\ge 1, \bigcap_{v\in {\mathscr N}_n} \{Z(v)=1\} \Big) 
=
\E\Big[ \Big( \frac{p}{p_c} \Big)^{\! N_n} {\bf 1}_{\{Y_n \ge 1\}}\Big],
$$

\noindent as desired.\qed 

\medskip
The following fact, which relies solely on the critical system $(Y_n)$, will be useful in proving Proposition \ref{p:lb}.

\begin{fact} [\cite{xyz_sustainability}]

 Assume \eqref{exp-assumption}. We have 
 \begin{equation}
     \P(N_n\ge 1)  = n^{-2 + o(1)}, \qquad n\to \infty. 
     \label{N>1}
 \end{equation}

 \noindent Fix any $\lambda>0$ and $\varrho>0$. For all sufficiently large $n$,
 \begin{equation}
     \P(Y_n \ge \ell+1, N_n \le   n^{2+\varrho})
     \ge
     n^{-(2+\varrho)}\,  m^{-\ell} ,
     \qquad
     \forall \ell \in [0, \, \lambda n]  .
     \label{YnNn}
\end{equation}
 
\end{fact}

\medskip

See \cite[Theorem 1.2 and Remark 2.4]{xyz_sustainability} for \eqref{N>1}, whereas \eqref{YnNn} is a straightforward consequence of \cite[Proposition 5.1 and (5.13)]{xyz_sustainability}.  

\subsection{Proof of Proposition \ref{p:lb}}

The main ingredient in the proof of Proposition \ref{p:lb} is the following deviation result for $N_n$, the number of open paths of the critical system $(Y_n)$.

\medskip

\begin{lemma} 
\label{p:lb_2}

 Assume \eqref{exp-assumption}. We have, for $j\to \infty$,
 \begin{equation}
     \label{smallN} 
     \liminf_{n\to\infty} \frac1{n} \log \P\Big( Y_n \ge \frac{n}{j}, \, N_n\le j n\Big) \ge - \frac{1}{j^{1 + o(1)}}, 
 \end{equation} 
 
 \noindent with $o(1) \to 0$ as $j \to \infty$. 

\end{lemma}

\begin{remark} 
\label{r:ldp} The correct rate in \eqref{smallN} should be $-\frac{O(1)}{j}$. Note that $ \P( Y_n \ge \frac{n}{j}, \, N_n\le j n) \le \P( Y_n \ge \frac{n}{j}) \le m^{-\frac{n}{j}}  \E (m^{Y_n}), $ and  $\E (m^{Y_n}) \to 1$ as $n \to \infty$ (see \cite[Theorem 3]{bmxyz_questions}), we obtain an upper bound for the probability term in \eqref{smallN}:  $ \limsup_{n\to\infty} \frac1{n} \log \P( Y_n \ge \frac{n}{j}, \, N_n\le j n) \le -\frac{\log m}{j}$.  It would be interesting to study the small deviation probabilities of $N_n$ conditioned  on survival: $\P(N_n \le \alpha_n \, |\, Y_n \ge 1)$ for $\alpha_n \to \infty$ and $\alpha_n=o(n^2)$. \qed

\end{remark}

\noindent {\it Proof of Lemma \ref{p:lb_2}.} The proof is based on an explicit construction of an event contained in $\{Y_n \ge \frac{n}{j}, N_n\le j\, n\}$. Let $\varrho \in (0, \, 1)$.  By \eqref{YnNn}, there exists $j_0 \ge 1$ such that for all $j\ge j_0$, 
\begin{equation} 
    \label{YjNj}
    \P \Big( Y_j \ge j+3, \, N_j \le j^{2+\varrho}\Big)  
    \ge 
    m^{-(j+2)} j^{-(2 +\varrho)}. 
\end{equation}

Fix $j\ge j_0$. For all $n>m j$, let $\ell = \ell(n, \, j) \ge 1$ be the smallest integer such that $n \le j + \ell +  j m^\ell$. Then $n > j+ \ell-1 + j m^{\ell-1}$, and $\ell \sim \frac{\log n}{\log m}$ as $n \to \infty$. Let $r=r(n, \ell):= \lceil \frac{n-j-\ell}{j} \rceil$ be  the smallest integer satisfying $r\ge \frac{n-j-\ell}{j}$. Then $m^{\ell-1} \le r \le m^{\ell}$.

We use the hierarchical construction presented in Section \ref{s:pre}. Recall \eqref{Yv}, \eqref{def-N}, and the fact that for any $|v|=j$, $(Y(v), N(v))$ is distributed as $(Y_j, N_j)$. By definition, 
$$
Y_{j+\ell} := Y({\mathfrak e}_{j+\ell}) \ge \sum_{|v|=j, \; v \in \T_{j+\ell}}  Y(v) - \sum_{k=0}^{\ell-1} m^k
\ge \sum_{v \in {\cal F}_r}  Y(v) - \sum_{k=0}^{\ell-1} m^k,
$$ 

\noindent where $\{|v|=j, \, v \in \T_{j+\ell}\}$ is the set of ancestors of  ${\mathfrak e}_{j+\ell}$ at generation $j$ (so its cardinality equals  $m^\ell$), and ${\cal F}_r$ denotes the first $r$ ancestors (in the lexicographic order) in this set. Let ${\cal F}^c_r:= \{|v|=j,\,  v \in \T_{j+\ell}\} \, \backslash {\cal F}_r$.  On the event $\cap_{v \in {\cal F}_r}\{Y(v)\ge j+3\}$, we have $Y_{j+\ell} \ge (j+3) r - \frac{m^\ell-1}{m-1} \ge (j+1) r$ as $\frac{m^\ell-1}{m-1} \le 2 m^{\ell-1} \le 2 r$. 

On the other hand, $N_{j+\ell} := N({\mathfrak e}_{j+\ell}) \le \sum_{|v|=j, \, v \in \T_{j+\ell}} N(v)$, so if $N(v) \le j^{2+\varrho}$ for all $v \in {\cal F}_r$ and $N(v)=0$ for all $v \in {\cal F}^c_r$, then $N_{j+\ell} \le r\, j^{2+\varrho} \le m^\ell j^{2+\varrho}$. Consequently,
\begin{eqnarray*}
 && \P\big( Y_{j+\ell} \ge (j+1) r, \, N_{j+\ell} \le m^\ell j^{2+\varrho}\big)
    \\
 &\ge& \P\Big(\bigcap_{v \in {\cal F}_r} \{Y(v)\ge j+3, N(v) \le j^{2+\varrho}\}\,  , \,  \bigcap_{v\in {\cal F}^c_r} \{N(v)=0\}\Big)
    \\
 &=& \P\big(Y_j\ge j+3, N_j \le j^{2+\varrho}\big)^r\, \P\big(N_j=0\big)^{m^\ell-r}
    \\
 &\ge& m^{-(j+2)r }\, j^{- (2+\varrho) r} \, \P\big( N_j=0 \big)^{m^\ell-r},
\end{eqnarray*}
 
\noindent where the last inequality follows from \eqref{YjNj}. By \eqref{N>1},  $\P\big(N_j=0\big) \ge \ee^{- j^{- 2 +\varrho}}$ for all $j\ge j_0$ (we may eventually enlarge the value of $j_0$ if needed), so $\P \big(N_j=0\big)^{m^\ell-r} \ge \P \big(N_j=0\big)^{m^\ell} \ge \ee^{- m^\ell j^{-2 +\varrho}}$. Thus we have proved that 
\begin{equation}
    \P\big( Y_{j+\ell} \ge (j+1) r, \, N_{j+\ell} \le m^\ell j^{2+\varrho}\big)
\ge m^{- j r }\, j^{- (2+\varrho) r} \, \ee^{- m^\ell j^{-2 +\varrho}}.
    \label{(Y,N)_at_level(j+ell)}
\end{equation}
 
Now we deal with $(Y_n, \, N_n)$. At generation $j+\ell$, we have a family $(Y(u), \, N(u))_{|u|=j+\ell} $ of $m^{n-(j+\ell)}$ i.i.d.\ copies of $(Y_{j+\ell}, \, N_{j+\ell})$. For $|u|=j+\ell$ with $u \in \T_n$, let $[u, \, {\mathfrak e}_n)$ be the path in $\T_n$ from $u$ to ${\mathfrak e}_n$ (including $u$ but excluding ${\mathfrak e}_n$); let 
$$
A_u:= \Big\{Y(u)\ge (j+1)r, \, N(u) \le m^\ell j^{2+\varrho}\Big\} \cap \bigcap_{w\in [u,\, {\mathfrak e}_n), \, v \in \mathtt{bro}(w)} \{N(v)=0\}.
$$

\noindent On $A_u$, we have $Y_n := Y({\mathfrak e}_n) = Y(u)- (n-j-\ell) \ge (j+1)r - (n-j-\ell) \ge r$ (the argument shows that $Y(w)>0$, a fortiori $N(w) >0$, for all $w\in [u,\, {\mathfrak e}_n]$: this property is going to be used in the next paragraph), and $N_n := N({\mathfrak e}_n) = N(u) \le m^\ell j^{2+\varrho}$. Therefore,
$$
\P(Y_n \ge r, \, N_n \le  m^\ell j^{2+\varrho})
\ge
\P \Big( \bigcup_{u \in \T_n: \; |u|=j+\ell} A_u \Big) \, .
$$

The events $A_u$ have the same probability. We claim that they are also disjoint. Indeed, if $A_u$ is realized for some $u$, then for all other $u' \in \T_n$ with $|u'|=j+\ell$, there exists $w\in [u,\, {\mathfrak e}_n)$ and $w'\in [u',\, {\mathfrak e}_n)$ such that $w'\in \mathtt{bro}(w)$. Thus $N(w')=0$ by definition of $A_u$. This implies that $A_{u'}$ is not realized (we have seen in the previous paragraph that on the event $A_{u'}$, we would have $N(w')>0$ for all $w'\in [u',\, {\mathfrak e}_n]$). We have thus proved that the events $A_u$ are disjoint. Consequently, 
$$
\P \Big( \bigcup_{u \in \T_n: \; |u|=j+\ell} A_u \Big)
=
m^{n-(j+\ell)} \, \P(A_{{\mathfrak e}_{j+\ell}}) \, ,
$$

\noindent which implies that
$$
\P(Y_n \ge r, \, N_n \le  m^\ell j^{2+\varrho})
\ge
m^{n-(j+\ell)} \, \P(A_{{\mathfrak e}_{j+\ell}}) \, .
$$

\noindent We observe that
$$
\P(A_{{\mathfrak e}_{j+\ell}})
=
\P \big( Y_{j+\ell } \ge (j+1)r, N_{j+\ell } \le m^\ell j^{2+\varrho}\big) \prod_{i=j+\ell}^{n-1} \P(N_i=0)^{m-1} \, .
$$

\noindent Consider the two probability expressions on the right-hand side. The first probability expression is at least $m^{- j r }\, j^{- (2+\varrho) r} \, \ee^{- m^\ell j^{-2 +\varrho}}$ (see \eqref{(Y,N)_at_level(j+ell)}), whereas $\P(N_i=0) \ge \ee^{-i^{-2+\varrho}}$ (see \eqref{N>1}). Therefore,
$$
\P(Y_n \ge r, \, N_n \le  m^\ell j^{2+\varrho})
\ge
m^{n-(j+\ell)- j r }\, j^{- (2+\varrho) r} \, \ee^{- m^\ell j^{-2 +\varrho}} \, \prod_{i=j+\ell}^{n-1} \ee^{-(m-1)i^{-2+\varrho} }.
$$
 
\noindent By the definition of $r$, we have $n-(j+\ell) - j r \ge -j$, and $j^{-(2+\varrho) r} \ge \ee^{-(2+\varrho) m^\ell \log j} \ge \ee^{- (2+\varrho) \frac{n m \log j}{j}}$. Note that  $m^\ell j^{-2+\varrho} \le n m j^{-3+\varrho}$ and $r\ge \frac{n}{2 j}$ for all large $n$, we get that 
$$
\P\Big( Y_n \ge \frac{n}{2 j}, \, 1 \le N_n \le  n m j^{1+\varrho} \Big) 
\ge 
\ee^{- 3m \frac{n \log j}{j}} ,
$$

\noindent for all large $n$. This proves Lemma \ref{p:lb_2} as $\varrho$ can be arbitrarily small.\qed
  
\bigskip

We have now all the ingredients for the proof of Proposition \ref{p:lb}.

\bigskip

\noindent {\it Proof of Proposition \ref{p:lb}.} By Theorem \ref{t:coupling}, for $p=p_c-\varepsilon$ with $\varepsilon\in (0, \, p_c)$,
\begin{eqnarray*}
    \P(X_n\ge 1)
 &\ge& \E\Big[ (1-p_c^{-1}\varepsilon)^{N_n} \, {\bf 1}_{\{ Y_n\ge 1, \, N_n\le \varepsilon^{-1/2}n\}} \Big]
    \\
 &\ge& (1-p_c^{-1}\varepsilon)^{ \varepsilon^{-1/2}n} \, \P( Y_n\ge 1, N_n\le \varepsilon^{-1/2}n), 
\end{eqnarray*}

\noindent which by \eqref{smallN} is larger than $(1-p_c^{-1}\varepsilon)^{ \varepsilon^{-1/2}n  }\, \exp ( -n \varepsilon^{\frac12 +o(1)})$ for all large $n$, with $o(1) \to 0$ as $\varepsilon\to 0$. This yields Proposition \ref{p:lb}. \qed

\section{Further remarks and questions}
\label{s:final}

We present some comments and questions to indicate a few important differences between the dual Derrida--Retaux conjecture studied in this paper and the (usual) Derrida--Retaux conjecture. 

\subsection{Existence of the limit}

It was explained in the introduction that part of the dual Derrida--Retaux conjecture was the existence of the constant
$$
\kappa(p) := \lim_{n\to \infty} \frac1n \log \E(X_n) ,
$$

\noindent for $p\in (0, \, p_c)$. We have not been able to prove the existence of the limit. For the (usual) Derrida--Retaux conjecture, the existence of the free energy
$$
F_\infty(p) := \lim_{n\to \infty} \frac{\E(X_n)}{m^n} ,
$$

\noindent is straightforward; indeed, we have seen in the introduction that $n\mapsto \frac{\E(X_n)}{m^n}$ is non-increasing.

\medskip

\begin{problem}

 Prove, under some suitable integrability assumption on the law of $X_0^*$, the existence of the limit
 $$
 \kappa(p) := \lim_{n\to \infty} \frac1n \log \E(X_n) ,
 $$

 \noindent for all $p\in (0, \, p_c)$.
 
\end{problem}

\subsection{``Mixing time"}

In the introduction, we explained the heuristics leading to the dual conjecture: if the initial distribution lies in an $\varepsilon$-neighbourhood of the critical manifold (meaning that $|p-p_c|$ is of order $\varepsilon$), then for a long time, of order $\varepsilon^{-1/2}$, the system lies in the $\varepsilon$-neighbourhood of the critical manifold before drifting away definitely. This phenomenon does not depend on the sign of $p-p_c$, and is common for both supercritical and subcritical regimes. 

As such, the time (of order $\varepsilon^{-1/2}$) during which the system lies in the $\varepsilon$-neighbourhood of the critical manifold before drifting away definitely plays a crucial role in both supercritical and subcritical regimes. In the supercritical regime, this time can be defined as the smallest integer $n$ such that $\E(X_n)$ exceeds $3$ (or any real number greater than $m^{1/(m-1)}$); the heuristics described in the previous paragraph can be made rigorous in a weaker form, which led to a proof of a weaker version of the Derrida--Retaux conjecture in \cite{bmvxyz_conjecture_DR}. In the subcritical regime, however, it is not clear how to define rigorously a quantity playing the role of this particular time (a kind of ``mixing time" necessary for a Markov chain to reach the stationary phase, except that here, it is the time necessary for the system to drift away), through the study of which one could prove the dual conjecture.

\medskip

\begin{problem}

 Define and study an appropriate ``mixing time" for the subcritical regime.
 
\end{problem}

\bigskip

\noindent {\Large\bf Acknowledgements} 

\medskip

\noindent We are grateful to Bernard Derrida for raising the problem and providing us with heuristics, and to Mikhail Lifshits for helpful discussions.

\end{document}